%% file: main.tex
\documentclass[preprint,twocolumn]{aastex631}

\usepackage{amsmath}

\begin{document}

\title{The Interplay of Parametric and Magnetorotational Instabilities in Oscillatory Shear Flows}
\shorttitle{PI vs MRI in Oscillatory Flows}
\shortauthors{Fairbairn \& Stone}

\author[0000-0002-9285-2184]{Callum W. Fairbairn}
\affiliation{Institute for Advanced Study \\
1 Einstein Drive, \\
Princeton NJ 08540, USA}

\author[0000-0001-5603-1832]{James M. Stone}
\affiliation{Institute for Advanced Study \\
1 Einstein Drive, \\
Princeton NJ 08540, USA}








\begin{abstract}
The evolution of warped disks is governed by internal, oscillatory shear flows driven by their distorted geometry. However, these flows are known to be vigorously unstable to a hydrodynamic parametric instability. In many warped systems, this might coexist and compete with the magnetorotational instability (MRI). The interplay of these phenomena and their combined impact on the internal flows has not been studied. To this end, we perform three-dimensional, magnetohydrodynamic unstratified shearing box simulations with an oscillatory radial forcing function to mimic the effects of a warped disk. In the hydrodynamic study, we find that the parametric instability manifests as strong, vertical `elevator' flows that resist the sloshing motion. Above a critical forcing amplitude, these also emerge in our magnetized runs and dominate the vertical stress, although they are partially weakened by the MRI, and hence the system equilibrates with larger radial sloshing flows. Below this critical forcing, the MRI effectively quenches the parametric instability. In all cases, we find that the internal stresses are anisotropic in character and better described by a viscoelastic relationship with the shearing flows. Unfortunately, these important effects are typically unresolved in global simulations of warped disks and are simplified in analytically tractable models. The incorporation of such complex, warp-amplitude-dependent, viscoelastic stresses will sensitively regulate the laminar flow response and inevitably modify the detailed spatio-temporal evolution of warped systems.
\end{abstract}



\input{0_introduction}
\input{1_problem_setup}
\input{2_parametric}

\input{3_hydro_slosh}

\input{4_mhd_slosh}
\input{5_discussion}

\input{6_conclusion}


\section*{Acknowledgments}

The authors thank the referee for a detailed and thought-provoking report. We would also like to thank Gordon Ogilvie, Geoffroy Lesur, and Martin Pessah for helpful conversations during the preparation of this manuscript. This research was supported by the W. M. Keck Foundation Fund and the IAS Fund for Memberships in the Natural Sciences (CWF).

%

\vspace{5mm}


\software{\texttt{AthenaK}: \cite{StoneEtAl_2024}}



\appendix
\input{app_parametric_instability}
\input{app_visc_hydro}

\bibliography{mybib}{}
\bibliographystyle{aasjournal}



\end{document}

%% file: 0_introduction.tex
\section{Introduction}
\label{sec:0_introduction}

Distorted astrophysical disks, which depart from the coplanar and circular assumptions of standard models, are now thought to be ubiquitous. Indeed, whenever there is some inherent misalignment present in a system, this can cause the disk to become warped, wherein the orbital plane varies as a function of the radius. For example, this misalignment might derive from the Lense-Thirring torque around tilted spinning black holes \citep[][]{BardeenPetterson_1975}, the inclined delivery of material during infall events \citep[][]{BateEtAl_2010,KuffmeierEtAl_2021}, the gravitational perturbations from tilted companions or planets \citep[][]{PapaloizouTerquem_1995,Xiang-GruessPapaloizou_2013,FacchiniEtAl_2013} or even inclined magnetic dipoles interacting with the surrounding disk \citep[][]{Lai_1999, RomanovaEtAl_2021}. This diversity of physical mechanisms is borne out in the varied observational evidence for warps. Warps have been indirectly inferred from the superorbital modulation of X-ray binary light sources \citep{Katz_1973, KotzeCharles_2012} and the shadows cast onto the outer regions of protoplanetary disks by inner warped regions \citep[][]{MarinoEtAl_2015,CasassusEtAl_2018}. Furthermore, warps have even been invoked to explain the variability of active galactic nuclei (AGN) \citep[][]{RajNixon_2021,KaazEtAl_2023} and the precession of jets \citep[][]{Miller-JonesEtAl_2019,CuiEtAl_2023a}. Tantalizing direct evidence for warps has also been provided by the warped maser emission in NGC 4258 \citep[][]{MiyoshiEtAl_1995} and the distorted midplane of the protostellar disks L1489 IRS \citep[][]{SaiEtAl_2020} and IRAS 04368+2557 \citep[][]{SakaiEtAl_2019}.

This prevalence of warped systems demands a rigorous theoretical understanding to model their evolution in space and time. Whilst the early studies laid much of the geometrical framework for global warp evolution \citep[see][]{Petterson_1977,Petterson1977b,Petterson_1978a}, they assumed that the disk behaves as a series of rigid, concentric annuli that interact in a purely viscous fashion. The paradigm was fundamentally changed by \cite{PapaloizouPringle_1983}, who realized that the distorted geometry actually establishes pressure gradients which drive internal, radial oscillatory shear flows that are odd about the midplane and efficiently communicate angular momentum. The pressure forces experienced by a fluid parcel vary over the orbital timescale as it traverses the warped geometry. This variation yields a near-resonant forcing, which matches the natural epicylic frequency. To regulate the resonant response, analytical work typically invokes the viscous $\alpha$ prescription \citep[][]{ShakuraSunyaev_1973} as a source of damping. Indeed, provided $\alpha$ exceeds the aspect ratio of the disk $H/R$, the viscosity effectively moderates the internal flows and sets up a quasi-steady `sloshing' state whilst the warp evolves over longer timescales -- this is the so-called \textit{diffusive} regime. In the opposite case that $\alpha < H/R$, the weaker viscosity is unable to temper the internal flows and instead the warp is detuned from resonance by dynamically evolving as a bending wave in the \textit{wavelike} regime \citep[][]{PapaloizouLin_1995}. These linear theories were subsequently extended toward nonlinear warp amplitudes by \cite{Ogilvie_1999} for the diffusive regime and by \cite{Ogilvie_2006, Ogilvie_2018} for the wavelike regime. Efforts to combine both regimes into a generalized set of evolutionary equations have also recently been pursued \citep[][]{MartinEtAl_2019,DullemondEtAl_2022a}.

All of these theories rely on the simplification that the disk flow structure is laminar and smooth and that any small-scale turbulence is well parameterized by the viscous $\alpha$ model. With this in mind, a number of global, laminar numerical experiments have found good agreement with the theoretical equations for warp evolution in both regimes \citep[e.g.][]{LodatoPrice_2010, FragnerNelson_2010, FacchiniEtAl_2013, NealonEtAl_2015, KimmigDullemond_2024a, FairbairnOgilvie_2023}. However, such a simple viscous closure model might not capture the true behavior of the underlying turbulence and its interplay with the warp. Indeed, it has long been appreciated that distorted disks are susceptible to a hydrodynamic, parametric instability which feeds off the internal, oscillatory shear flows \citep[][]{PapaloizouTerquem_1995,GammieEtAl_2000}. This small-scale instability is most readily understood within a local warped shearing box model \citep[][]{OgilvieLatter2013a,OgilvieLatter2013b} which was numerically implemented in the two-dimensional simulations of \cite{PaardekooperOgilvie_2019}. They recovered the linear growth phase of the instability and the ensuing nonlinear saturated state. This significantly modified the laminar flows and reduced the associated angular momentum transport in the warped disk. Furthermore, they found complicated variability in the final turbulent state, as a function of the warp amplitude, and also evidence of hysteresis as the saturated flows depended on the warp history. This instability can also efficiently extract energy from the sloshing flows and lead to a rapid damping of the driving warp \citep[][]{FairbairnOgilvie_2023}. Unfortunately, this richness of dynamics is typically unresolved in global numerical experiments. The only clear identification so far has been the high-resolution, 120 million Lagrangian particle runs of \cite{DengEtAl_2020} and \cite{DengOgilvie_2022a}, which attains $\sim 8$ elements per scale height. The emergence of the parametric instability dramatically damps the warp, strongly departing from the laminar prescriptions.

This purely hydrodynamic picture is further complicated by the possible presence of magnetic fields in warped disks -- for example, relevant to the case of X-ray binaries or AGN disks. Analytical progress on this front is limited to the regime of strong magnetization for which vertical field lines threading the disk provide an effective tension -- resisting the oscillatory, shearing flows and detuning the aforementioned Keplerian resonance \citep[][]{ParisOgilvie_2018}. In the alternate case of weak fields, the disk is readily unstable to the magnetorotational instability (MRI) \citep{BalbusHawley_1991,HawleyBalbus_1991}. Whilst this mechanism is often invoked to explain radial transport of mass and angular momentum, the turbulence might also help moderate the vertically shearing, sloshing flows. Indeed, often the in-plane accretion stresses are equated to some value of effective $\alpha$, which is then used in the isotropic, viscous theories outlined above. In reality, the picture is inevitably more complicated as the MRI turbulence is anisotropic and perhaps does not respond instantaneously to the local shear rate. Therefore, we need to appeal to numerical simulations to gain further insight. Indeed, the global, magnetized warped torus simulations of \cite{SorathiaEtAl_2013b} suggested a distinct evolution from that predicted by viscous theory. Furthermore, they found that the flow was dominated by hydrodynamic rather than magnetized stresses, akin to their purely hydrodynamic runs in \cite{SorathiaEtAl_2013a}. Contrastingly, \cite{NealonEtAl_2016} were able to reproduce the MHD simulations of a warped disk around a spinning black hole \citep[][]{KrolikHawley_2015} using viscous simulations. Despite these global efforts, there has been little work to investigate in detail how the local internal flows are modified by the MRI. Exceptionally, \cite{TorkelssonEtAl_2000} performed three-dimensional stratified, shearing box simulations of an MRI turbulent disk where the effects of the warp are mocked up via an imposed epicylic, sloshing initial condition that is odd about the midplane. This controlled experiment identified the early emergence of the parametric instability, followed by exponential damping of the sloshing motions due primarily to the hydrodynamic transport of momentum. The damping timescales found are consistent with vertical viscosities which are (coincidentally) comparable to the in-plane MRI effective viscosity -- thus supporting an isotropic $\alpha$.

Despite this progress, there are several routes demanding further inquiry. The MRI simulations of \cite{TorkelssonEtAl_2000} implement the warp as a free initial condition that quickly decays. Therefore, there is little time to establish a saturated, magnetized turbulent state in which the parametric instability, MRI, and sloshing flows come into an equilibrium. Indeed, extracting the nature of the turbulent stresses is complicated by the limited timescales. Contrastingly, the detailed parametric instability study of \cite{PaardekooperOgilvie_2019} establish steady sloshing states but focus on two-dimensional hydrodynamics. In this paper, we will find a compromise between the two approaches. In particular, we will compare controlled hydrodynamic and magnetized, local simulations of oscillatory shear flows that are maintained by a time-dependent forcing function, mimicking the radial pressure gradients set up by warped disks. In this initial study, we will further simplify matters by considering the unstratified problem, which will be extended toward the stratified case in future work. The outline for this paper is as follows. In Section \ref{sec:problem_setup} we will describe the problem setup and summarize the laminar model predictions. In Section \ref{sec:parametric}, we will benchmark our numerical setup by capturing the linear growth of the parametric instability and compare it with theoretical predictions. In Section \ref{sec:hydro_slosh}, we will examine the fiducial hydrodynamic sloshing run and study the emerging stresses. This will be compared with the fiducial magnetized sloshing run in Section \ref{sec:mhd_slosh}, before we perform a study varying the forcing amplitude. We will discuss our findings in the context of previous studies in Section \ref{sec:discussion} and finally conclude in Section \ref{sec:conclusion}.

%% file: 1_problem_setup.tex
\section{Problem Setup}
\label{sec:problem_setup}

In this section, we will outline the general setup considered throughout this paper. Since we are investigating the zoomed-in dynamics of distorted flows, we naturally appeal to the local shearing box approximation \citep[][]{Hill_1878,HawleyEtAl_1995} wherein the fluid equations are locally expanded about a reference orbit at radius $r_0$, with angular velocity $\Omega_0(r_0)$. This reference frame is furnished with a corotating, cartesian coordinate system $(x,y,z)$ which is related to the standard cylindrical coordinates via $x = (r-r_0)$, $y = r_0(\phi-\Omega_0 t)$ and $z = z$. Within this local expansion, the governing magnetohydrodynamic (MHD) equations in conservative form are written 
\begin{eqnarray}
    \label{eq:continuity}
    && \frac{\partial \rho}{\partial t}+\nabla\cdot(\rho \mathbf{u}) = 0, \\
    \label{eq:momentum}
    && \frac{\partial (\rho \mathbf{u})}{\partial t}+\nabla\cdot\left[\rho\mathbf{u}\mathbf{u}-\mathbf{B}\mathbf{B}\right]+\nabla P^* = \nonumber\\ 
    && -2\Omega_0\hat{\mathbf{z}}\times\rho\mathbf{u}+2\rho\Omega_0^2qx\hat{\mathbf{x}}+\nabla\cdot\mathsf{T} , \\
    \label{eq:induction}
    && \frac{\partial B}{\partial t} - \nabla\times\left(\mathbf{u}\times\mathbf{B}-\eta\nabla\times\mathbf{B}\right) = 0,
\end{eqnarray}
where $\rho$ is the density, $\mathbf{u}$ is the velocity vector, and $\mathbf{B}$ is the magnetic field. The total pressure $P^{*} = p+\mathbf{B}\cdot\mathbf{B}/2$, is composed of the magnetic pressure and the thermal pressure $p$. Throughout this work, we will assume the simplest isothermal equation of state (EoS) $p = c^2\rho$, connecting pressure and density via the sound speed $c$. This removes the need to explicitly solve an energy equation. On the right-hand side of equation \eqref{eq:momentum}, we see contributions from the Coriolis force and the tidal potential, featuring the shear parameter $q = -d\ln\Omega/d\ln r = 3/2$, for a Keplerian disk. Furthermore, the shear viscous stress tensor \citep[][]{LandauLifshitz_1959} is defined as
\begin{equation}
    \label{eq:viscous_stress}
    \mathsf{T}_{ij} = \rho\nu_{ij}\left(\frac{\partial u_i}{\partial x_j}+\frac{\partial u_j}{\partial x_i}-\frac{2}{3}\delta_{ij}\nabla\cdot\mathbf{u} \right) ,
\end{equation}
where we allow for an anisotropic kinematic viscosity $\nu_{ij}$.\footnote{Note that symmetry imposes $\nu_{ij} = \nu_{ji}$. Furthermore, this simple prescription for characterizing anisotropy means that $\mathsf{T}_{ij}$ does not generally transform in a consistent tensorial fashion.} Another nonideal effect is included in the induction equation \eqref{eq:induction} when the ohmic resistivity $\eta$ is set to be nonzero. Note that for the purposes of this focused study, we have neglected the vertical component of gravity and therefore consider unstratified disk setups. This extension will be addressed in future work.  

\begin{figure*}
    \centering
    \includegraphics[width=\linewidth]{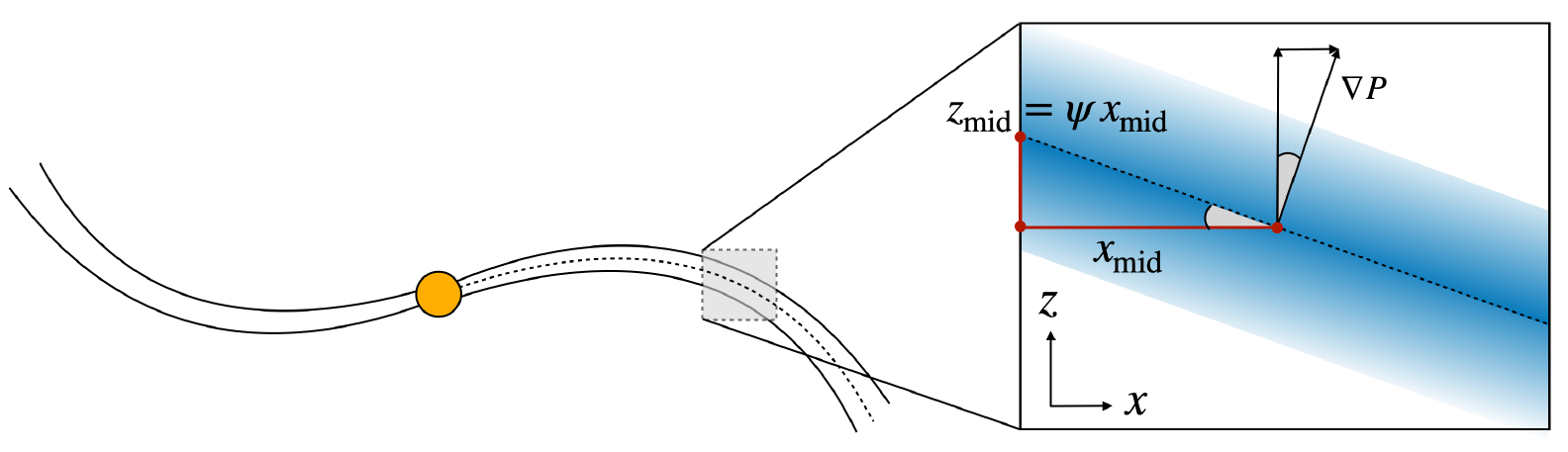}
    \caption{Global and local cartoon of a warped disk. \textit{Left}: The global warped disk consists of a series of orbits with a radially varying orientation of angular momentum. \textit{Right}: A zoomed-in patch of the warped disk manifests as a tilting of the midplane. This projects a component of the vertical pressure gradient in the radial direction and drives sloshing flows.}
    \label{fig:figure_global_local}
\end{figure*}

This study is primarily interested in probing the effect of instabilities and ensuing nonlinear turbulence in moderating oscillatory flows forced by distorted geometries. In particular, it has long been appreciated that warped disk evolution is driven by self-induced internal flows \citep[e.g.][]{PapaloizouPringle_1983,OgilvieLatter2013a}. To gain some heuristic intuition, consider the warped disk cartoon in Fig.~\ref{fig:figure_global_local}. The globally distorted geometry is visualized on the left, where the midplane changes inclination as a function of the radius. The orientation of the angular momentum on each disk annulus is formally encoded by the radial and temporal variation of the unit tilt vector $\hat{\mathbf{l}}(r,t)$. The warp amplitude is then quantified by $\psi = \left|d\hat{\mathbf{l}}/d\ln r\right|$. On the right, we zoom in to a portion of the disk and examine the local manifestation of the warp as a tilted midplane, defined by $z_{\rm mid} = \psi x_{\rm mid}$. In a flat, stratified disk, the vertical hydrostatic pressure gradient is given by $\partial p/\partial z = -\rho\Omega_0^2z$. Therefore, upon tilting the midplane, a portion of this force is projected into the radial direction, yielding $-\mathbf{\hat{x}}\cdot\nabla p\sim\rho\Omega_0^2\psi z$ \citep[][]{LodatoPringle_2007}. As we track the orbit around the disk, the midplane tilts back and forth, incurring a sinusoidal dependence in the radial forcing such that it has the form $f_{\rm w}\sim \rho\Omega_0^2\psi z \cos(\Omega_0 t)$. The characteristic properties of this forcing are that it is linear in the vertical coordinate $z$ and oscillates on the orbital timescale. 

Of course, the simple, unstratified shearing box setup introduced above precludes the self-consistent generation of these pressure gradients. Instead, we will implement an artificial forcing function that emulates key properties in a controlled fashion. To this end, we will add
\begin{equation}
    \label{eq:warp_forcing}
    \mathbf{f}_{\rm w} =  \mathrm{Re}\left[\rho \Omega_0^2 H \psi_{\rm f} \sin\left(\frac{2\pi z}{L_z}\right)\exp(i\Omega_0 t)\hat{\mathbf{x}}\right],
\end{equation}
to the right-hand side of equation \eqref{eq:momentum}, where $\psi_{\rm f}$ is now a complex number allowing for an arbitrary choice of phase. Here, $L_z$ is the vertical domain in our shearing box, whilst $H \equiv c/\Omega_0$ sets the characteristic scale height. Near the box midplane, this form approximates a linear function of $z$ before turning around to satisfy periodic boundary conditions. This idealized model, although simple, contains all of the fundamental physics for driving sloshing flows in warped disks. An alternative approach might be to use the warped generalization of the shearing box, developed by \cite{OgilvieLatter2013a}, which explicitly incorporates the warped geometry via Lagrangian coordinates that track the tilted orbital motion. However, in order to leverage current state-of-the-art MHD numerical code bases, with inbuilt support for standard shearing boxes, we will adopt our simplified approach. 

To gain some intuition about the growth of sloshing motions in our setup, consider the hydrodynamic limit of equations \eqref{eq:continuity} and \eqref{eq:momentum}, equipped with the forcing function in equation \eqref{eq:warp_forcing}. Furthermore, we will assume that the flow structure is horizontally homogeneous, laminar, and moderated by the shear viscosity, which acts as a model for the turbulent stresses investigated later in this paper. Motivated by the sinusoidal vertical structure of $\mathbf{f}_{\rm w}$ the velocity response follows
\begin{eqnarray}
    \label{eq:laminar_ansatz_A}
    && u_{x,{\rm w}} = A(t) c \sin(k_{\rm b} z) , \\
    \label{eq:laminar_ansatz_B}
    && u_{y,{\rm w}} = u_{\rm K} + B(t) c \sin(k_{\rm b} z),
\end{eqnarray}
where $u_{\rm K} = -q\Omega_0 x$ is the usual Keplerian shear flow solution in the absence of warped forcing, and we have defined $k_{\rm b} \equiv 2\pi/L_z$. It is also sometimes useful to work with the velocity relative to the background Keplerian flow, which we will denote $\mathbf{v} = \mathbf{u}-\mathbf{u}_{\rm k}$. Inserting this ansatz into the governing equations yields
\begin{eqnarray}
    \label{eq:Adot_eqn}
    && \dot{A}-2\Omega_0 B = \Omega_0 \psi_{\rm f} e^{i\Omega_0 t} - \nu_{xz} k_{\rm b}^2 A, \\
    \label{eq:Bdot_eqn}
    && \dot{B}+\Omega_0 A/2 = -\nu_{yz} k_{\rm b}^2 B,
\end{eqnarray}
which can be combined into the forced, damped simple harmonic oscillator equation\footnote{Here we are working with full complex generality, although it should be remembered that in practice we actually extract the real parts of $A$ and $B$.}
\begin{eqnarray}
    \label{eq:forced_sloshing_SHO}
    \ddot{A}+(\nu_{xz}+\nu_{yz})k_{\rm b}^2 \dot{A}+(\Omega_0^2+\nu_{xz}\nu_{yz}k_{\rm b}^4)A = \nonumber\\
    (i\Omega_0^2+\nu_{yz}k_{\rm b}^2 \Omega_0)\psi_{\rm f} e^{i\Omega_0 t}.
\end{eqnarray}
In the simplest case for which the system is unforced and inviscid, the dynamics reduces to the epicylic oscillator solution 
\begin{equation}
    \label{eq:epicylic_slosh}
    \begin{pmatrix}
    A \\
    B
    \end{pmatrix} = 
    S\begin{pmatrix}
        1 \\
        i/2
    \end{pmatrix} e^{i\Omega_0 t},
\end{equation}
for which $S$ sets the sloshing amplitude and phase. Note that equation \eqref{eq:epicylic_slosh} represents an exact solution to the (unforced) nonlinear shearing box equations, absent magnetic fields. The vertical layers oscillate independently as a series of stacked epicylic orbits. The amplitude of these oscillations as a function of $z$ is enforced by our ansatz in equations \eqref{eq:laminar_ansatz_A} and \eqref{eq:laminar_ansatz_B}, although for this free sloshing, any arbitrary function of $z$ is obviously permitted. We will restrict ourselves to this coherent sinusoidal structure, which is established by our forcing function. A useful diagnostic which encodes the magnitude of this $x$--$y$ sloshing motion is the epicylic amplitude
\begin{equation}
    C = A^2+4B^2.
\end{equation}
This is clearly a conserved quantity for the free sloshing motion when equation \eqref{eq:epicylic_slosh} is inserted.

When the system is inviscid but now forced, equation \eqref{eq:forced_sloshing_SHO} reduces to the standard, resonantly forced oscillator with a linearly growing response
\begin{eqnarray}
    \label{eq:resonant_response_A}
    && A_{\rm res}(t) = \frac{\psi_{\rm f}}{2}\left(\Omega_0 t -i\right)e^{i\Omega_0 t} , \\
    \label{eq:resonant_response_B}
    && B_{\rm res}(t) = \frac{\psi_{\rm f}}{4}i\Omega_0 te^{i\Omega_0 t} .
\end{eqnarray}
This resonance between the warp-induced radial forcing frequency and the orbital frequency (equal to the natural epicyclic frequency for our Keplerian shear) underlies the difficulty in developing a theory for warps in the low viscosity, bending wave regime, where the response is dynamical and no quasi-steady sloshing state can be found. Meanwhile, the inclusion of viscosity tempers the resonance and gives the steady-state sloshing
\begin{eqnarray}
    \label{eq:viscous_response_A}
    && A_{\rm vis}(t) = \frac{(\nu_{yz}k_{\rm b}^2 +i\Omega_0)\Omega_0\psi_{\rm f}}{\nu_{xz}\nu_{yz}k_{\rm b}^4+i\Omega_0(\nu_{xz}+\nu_{yz})k_{\rm b}^2}e^{i\Omega_0 t} , \\
    \label{eq:viscous_response_B}
    && B_{\rm vis}(t) = -\frac{\Omega_0^2\psi_{\rm f}}{2\left[\nu_{xz}\nu_{yz}k_{\rm b}^4+i\Omega_0(\nu_{xz}+\nu_{yz})k_{\rm b}^2\right]}e^{i\Omega_0 t}.
\end{eqnarray}
Adopting the simplification that $\nu_{xz} = \nu_{yz} = \nu$ and taking the limit $\nu k_{\rm b}^2 \ll \Omega_0$ yields the amplitude
\begin{equation}
    \label{eq:simple_alpha}
    |A_{\rm vis}| \sim \frac{\Omega_0\psi_{\rm f}}{2\nu k_{\rm b}^2}.
\end{equation}
Therefore, given the forcing amplitude and by measuring the steady sloshing response, we can extract the viscosity. This procedure will allow us to relate the realistic turbulence, emerging in the simulations of sections \ref{sec:hydro_slosh} and \ref{sec:mhd_slosh}, to an effective viscous laminar model.

%% file: 2_parametric.tex
\section{Parametric Instability}
\label{sec:parametric}

In order to benchmark the numerical implementation of the setup described in Section \ref{sec:problem_setup}, it is helpful to first compare with a controlled problem for which we have some analytical insight. Therefore, we will start by considering the stability of a low-amplitude, inviscid, free sloshing flow (as described by equations \eqref{eq:laminar_ansatz_A}, \eqref{eq:laminar_ansatz_B} and \eqref{eq:epicylic_slosh}), which is imposed as an initial condition. 

\subsection{Summary of the Weakly Nonlinear Parametric Instability}
\label{subsec:WN_PI}

Such sloshing flows are known to be unstable due to the parametric instability (PI) \citep[e.g.][]{GammieEtAl_2000}, which feeds off the time-dependent, shearing flow. When the shear flow is sufficiently subsonic, with $S \ll 1$, the mechanism can be understood as a three-mode coupling, whereby the sloshing motion acts as the parent mode which destabilizes a pair of inertial waves. The detailed weakly nonlinear theory of this process is expounded in appendix \ref{app:wn_parametric_instability}, adapting the analysis of \cite{OgilvieLatter2013b} and \cite{FairbairnOgilvie_2023} to our present setup. Here, we will simply summarize the main results. 
\begin{figure}
    \centering
    \includegraphics[width=\linewidth]{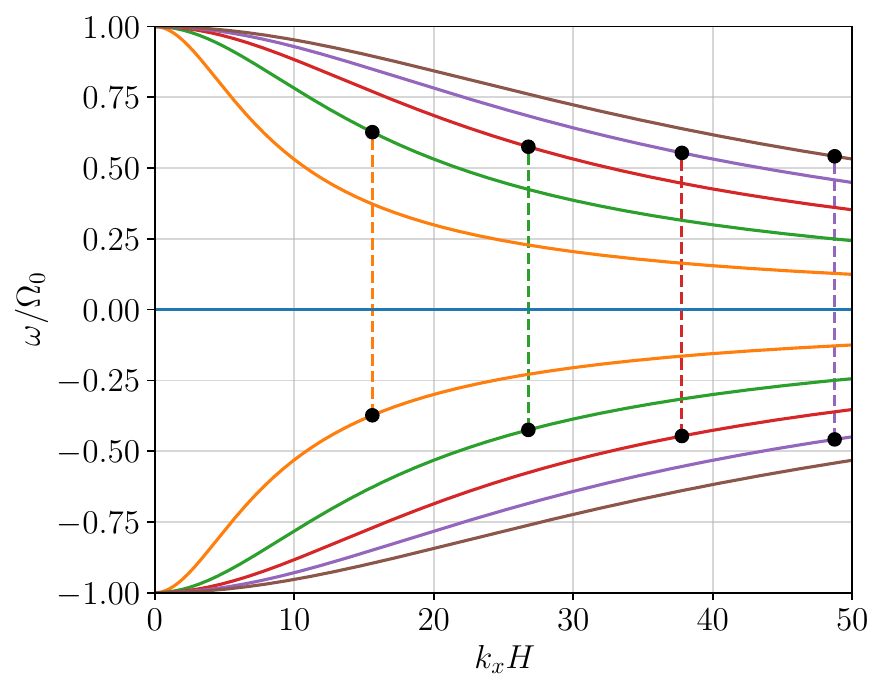}
    \caption{Dispersion relation curves tracing the frequency vs radial wavenumber relation for locally axisymmetric inertial waves supported by our unstratified shearing box. The colored lines denote different branches, labeled by the vertical modal order $n$ (0: blue, 1: orange, 2: green, 3: red, 4: purple, 5: brown). The connective dashed lines indicate the resonant couplings susceptible to parametric instability.}
    \label{fig:figure_dispersion_relation}
\end{figure}

The shearing box supports a variety of wave modes, wherein perturbed variables follow the modal ansatz $X^\prime = X_n\exp[i(k_x x+k_{\rm b} n z-\omega t )]$. When $S = 0$, these modes are decoupled and follow a dispersion relation (described by equation \eqref{eq:dispersion_relation}). This is visualized for inertial waves in Fig.~\ref{fig:figure_dispersion_relation}. The different vertical orders are traced out by the family of colored lines. For example, the orange line denotes $n=1$, which jumps to higher $n$ branches with increasing frequencies, for a given $k_x$. However, the introduction of sloshing facilitates a coupling between modes that can lead to instability. The dashed lines indicate mode couplings that resonate effectively with the background sloshing motion. This requires that the inertial waves be of neighboring order in $n$ and are separated in frequency by $\Omega_0$, the sloshing frequency
as seen in the coorbital frame. When these conditions are satisfied, energy from the shear flow is efficiently  pumped into the inertial waves, which grow according to 
\begin{equation}
    \label{eq:PI_growth_rate}
    \frac{\sigma}{\Omega_0} = S\sqrt{C_1 C_2},
\end{equation}
where $C_1$ and $C_2$ are defined by equations \eqref{eq:C1}-\eqref{eq:C2} and depend on the pair of inertial modes involved.

\subsection{Floquet Theory}
\label{subsec:floquet}

The predictions of the weakly nonlinear theory can be independently verified by appealing to Floquet theory. Indeed, the linearized equations \eqref{eq:non_dimensional_un}-\eqref{eq:non_dimensional_hn}, which govern the stability of the sloshing flow, can be written in the form 
\begin{equation}
    \label{eq:floquet_form}
    \dot{\mathbf{V}} = \mathsf{N}(t)\mathbf{V},
\end{equation}
where $\mathbf{V}$ is a vector of length $4(2N_{\rm max}+1)$ containing the modal, perturbation coefficients $[u_n, v_n, w_n, h_n]$, truncated at some order in $|n|\leq N_{\rm max}$. Meanwhile, $\mathsf{N(t)}$ is the matrix encoding the ordinary differential equation coefficients, which are periodic over the dynamical timescale $T = 2\pi/\Omega_0$. We integrate this equation for each of the unit basis vectors, spanning all possible initial conditions for $\mathbf{V}(0)$. The resulting vectors after one sloshing period, $\mathbf{V}(T)$, can be arranged as the columns of the monodromy matrix $\mathsf{M}(T)$. The eigenvalues of this matrix are the characteristic multipliers $e^{\sigma T}$, whilst the associated eigenvectors $\mathbf{V}_{\rm f}$ define the modal initial conditions which evolve according to each multiplier. In particular, there exists a solution to equation \eqref{eq:floquet_form}, which evolves according to $\mathbf{V}(t+T) = \mathbf{V}(t) e^{\sigma T}$ where $\mathbf{V}$ is periodic over $T$ and $\mathbf{V}(0) = \mathbf{V}_{\rm f}$. 

\begin{figure}
    \centering
    \includegraphics[width=\linewidth]{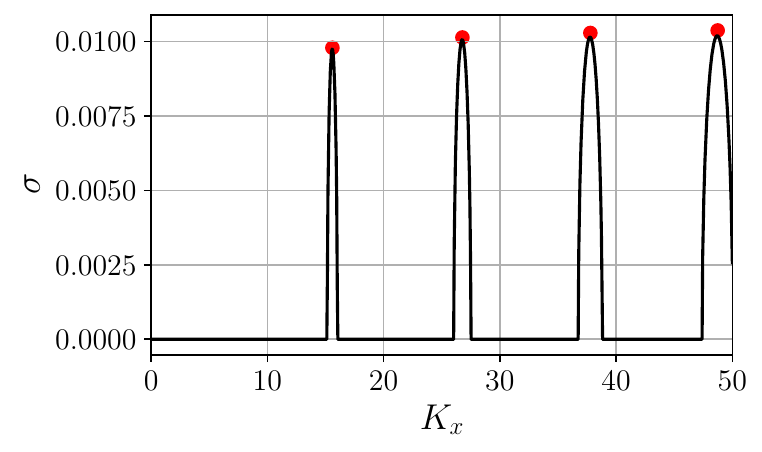}
    \caption{Maximal growth rates $\sigma$, obtained via the Floquet method (\textit{black line}) vs three-mode coupling theory (\textit{red markers}) for $S = 0.01$ across a range of $k_x$.}
    \label{fig:figure_floquet}
\end{figure}

Let us test this methodology against the predictions of Section \ref{subsec:WN_PI}. For the applicability of weakly nonlinear theory, we require a small sloshing amplitude and thus adopt $S = 0.01$. We then truncate our system of Floquet equations to modes with $|n|\leq 20$ and solve for the monodromy matrix across a range of $k_x$. For each wavenumber, we extract the maximum characteristic multiplier and associated growth rate, $\sigma$. The results are plotted in Fig.~\ref{fig:figure_floquet}. The black lines show the results from the Floquet analysis, which exhibit a series of unstable bands that coincide with the resonant wavelengths predicted by the dashed lines in Fig.~\ref{fig:figure_dispersion_relation}. Meanwhile, the red markers plot the growth rates predicted from equation \eqref{eq:PI_growth_rate} of the three-mode coupling theory. Clearly, there is very good agreement in this weak sloshing regime, mutually verifying our analysis and the Floquet implementation. We have also tested (not shown here) that the Floquet growth rates vary linearly in $S$, in accordance with theory, provided that the flows remain in the weak sloshing regime. Nonetheless, it should be emphasized that the applicability of Floquet theory extends beyond this weakly nonlinear regime, since the problem structure follows equation \eqref{eq:floquet_form} for arbitrary values of $S$. In fact, this will provide a useful tool to compare with our simulations in sections \ref{sec:hydro_slosh} and \ref{sec:mhd_slosh}, where the sloshing magnitude can become of sonic order. However, at such a large $S$, one should be careful to include sufficient modes in the Floquet analysis to ensure convergence. Indeed, the strong sloshing facilitates enhanced modal communication, which will spread the couplings beyond three waves. Furthermore, there is the possibility that there are no modal solutions for very large shear flows, as discussed by \cite{OgilvieLatter2013b} and \cite{PaardekooperOgilvie_2019}, which would preclude convergence entirely.

\subsection{Numerical Verification}
\label{subsec:PI_numerical_tests}

We set up this free sloshing test problem in \texttt{AthenaK} -- a highly portable \texttt{Kokkos} implementation of \texttt{Athena++}, which can be flexibly deployed on a variety of architectures \citep[][]{StoneEtAl_2024}. Throughout this paper, \texttt{AthenaK} will be our tool of choice to solve the MHD equations \eqref{eq:continuity}-\eqref{eq:induction}. To this end, we employ the three-dimensional shearing box module, as described by \cite{StoneGardiner_2010a}, equipped with radial shearing periodic boundary conditions and $q = 3/2$. Furthermore, we will adopt periodic boundary conditions along the $y$ and $z$ dimensions. Presently, we study the ideal hydrodynamic problem and switch off the viscosity and magnetic fields. We adopt code units such that $c = H = \Omega_0 = 1$, while the initialized uniform density profile is freely set as $\rho_0 = 1$. Atop the standard, Keplerian shearing background, we add the radial sloshing initial condition with $u_x = S c \sin{(2\pi z/L_z)}$. We choose $S = 0.01 \ll 1$ so as to reside in the weakly nonlinear regime which interfaces with the theory described in Section \ref{subsec:WN_PI}.

We choose our box size judiciously so that it permits the unstable radial wavenumbers predicted from the three-mode coupling theory. In particular, we will test the growth of the longest radial wavelength resonance, as shown by the dashed orange line in Fig.~\ref{fig:figure_dispersion_relation}, which occurs at $k_x H = 15.6$ and connects the $n=1$ branch at $\omega/\Omega_0 = -0.373$, with the $n=2$ branch at $\omega/\Omega_0 = 0.627$. In order to allow for three radial wavelengths to fit in the box, we set $L_x = 1.2084$. Meanwhile, we choose $L_y = \pi$ and $L_z = H = 1$. Capturing the growth rate accurately requires a sufficient resolution, so we establish a grid with $(N_x,N_y,N_z) = (154,200,128)$ which well resolves the vertical and radial wavelengths. We then seed the instability by adding the resonant pair of inertial modes, each with eigenmode structure $\mathbf{V}_n^{(0)}(\omega;k_x H)$ described by equation \eqref{eq:eigenvector}. Each mode is normalized so that the maximal radial velocity associated with it is 0.0001 -- a factor of 100 smaller than the sloshing. We then run the simulation on 1 H200 GPU for 60 orbital periods.
\begin{figure}
    \centering
    \includegraphics[width=\linewidth]{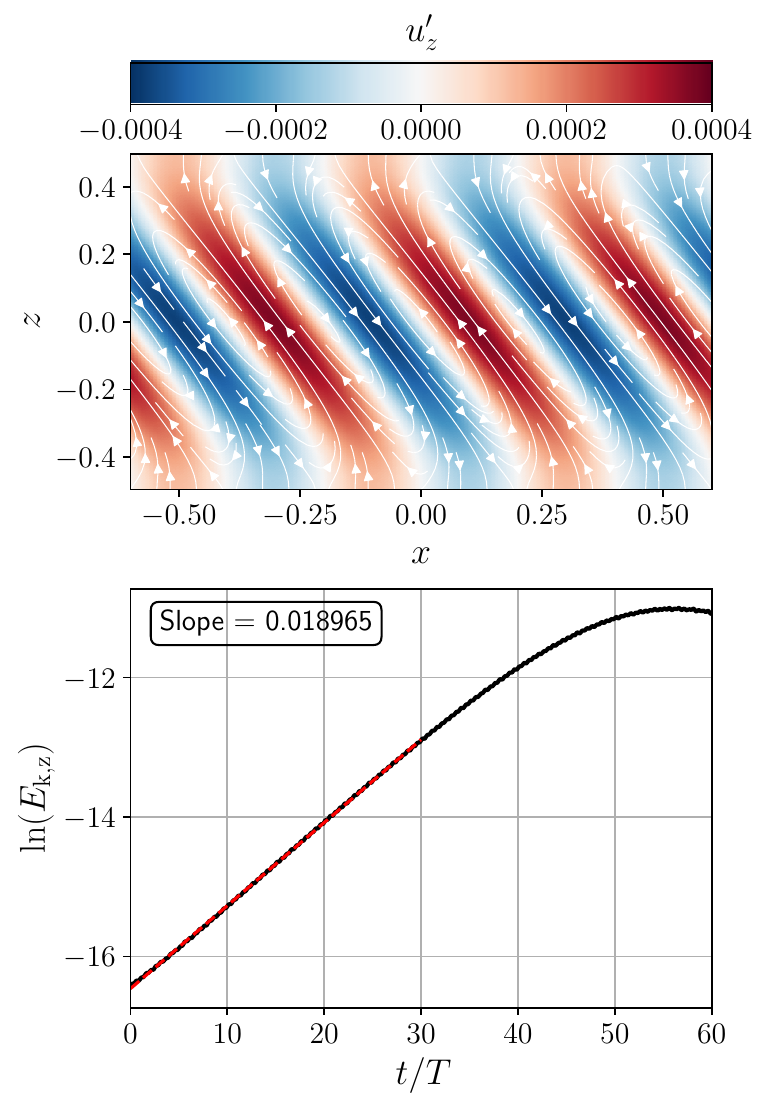}
    \caption{Numerical simulation verifying three-mode coupling parametric instability. \textit{Upper panel}: The initialized modal structure in $u_z^\prime$ is shown as the heat map, overplotted with white arrows denoting the meridional streamlines. \textit{Lower panel}: The growth of the vertical kinetic energy in our simulation (black line) is fit (red dashed line) between $0<t<30T$ in order to extract the energy growth rate $2\sigma = 0.018965$. }
    \label{fig:figure_PI_mode_growth}
\end{figure}
As desired, our numerical experiment nicely captures the modal amplification by parametric instability. The key results are visualized in Fig.~\ref{fig:figure_PI_mode_growth}. The heat map in the upper panel shows the initialized modal structure of the vertical velocity perturbation $u_z^\prime$. Meanwhile, the meridional streamlines are overplotted as the white arrows. These oblique vertical flow structures are also typical signatures of the PI, which manifest in other disk setups \cite[e.g.][]{OgilvieLatter2013b,DengEtAl_2020}. In the lower panel, the black line plots the logarithm of the total vertical component of the kinetic energy $E_{{\rm k}, z}$ summed over the box. This undergoes a clear exponential rise phase, which is well fit by the red dashed line between $0<t<30T$. The slope of this line indicates an energy growth rate of $0.018965$, which yields an amplitude growth rate of $\sigma = 0.0094825$. Pleasingly, this is very close to the predictions of the three-mode coupling theory, which gives $\sigma_{\rm WN} = 0.00978919$ (as plotted by the first red marker in Fig.~\ref{fig:figure_floquet}). One might rush to a conclusion that the slight discrepancy here is due to numerical dissipation. However, we have also performed tests with double and quadrupole the grid resolution and confirm that our fiducial setup is already suitably converged. Indeed, if we initialize a free wave and switch off the free sloshing, we can track its numerical dissipation, which is found to have an energy damping rate of $\sim 0.0001< |\sigma_{\rm WN}-\sigma|$. This is too small to explain the aforementioned difference between the predicted and measured growth rates. Instead, the slight discrepancy is mainly attributed to subdominant, direct nonlinear interactions between the two waves, which are neglected in the weakly nonlinear analysis. Indeed, the smooth turnover in $\ln(E_{\rm k,z})$ around $t = 50T$ is also indicative of these nonlinear, secular effects. Additional tests, using waves initialized with smaller seed amplitudes, show this variation over longer timescales, and the growth rates converge toward the weakly nonlinear predictions.

%% file: 3_hydro_slosh.tex
\section{Hydrodynamic results}
\label{sec:hydro_slosh}

Having outlined the theoretical setup in Section \ref{sec:problem_setup}, and benchmarked the numerics with a controlled problem in Section \ref{sec:parametric}, we are now in a position to examine the role of hydrodynamic instability and the ensuing turbulence in setting an equilibrium state when the sloshing is forced.

\subsection{Numerical Setup}
\label{subsec:hydro_numerical_setup}

The numerical setup is similar to that described in Section \ref{subsec:PI_numerical_tests}, but now we forgo the free sloshing initial condition (setting $S = 0$) and instead employ the forcing function described by equation \eqref{eq:warp_forcing} as a user-implemented source term in \texttt{AthenaK}. Our fiducial study will adopt a value of $\psi_{\rm f} = 0.3$ which is large enough to set up a clear sloshing signal atop the turbulence. We will compare this with smaller forcing values in Section \ref{subsec:forcing_amplitude}. Once again, we set the explicit viscosity to be 0 for these hydrodynamic runs. In appendix \ref{app:visc_hydro} we also perform a number of viscous experiments, for comparison with our MHD runs (see Section \ref{sec:mhd_slosh}) which necessarily include nonideal dissipation in order to obtain a converged MRI. Our box size is taken to be $(L_x,L_y,L_z) = (9.6672,\pi,1)$, which is elongated by a factor of 8 in the $x$ dimension compared to the experiment performed in Section \ref{subsec:PI_numerical_tests}. As we will show in the following results, the forced sloshing, turbulent state wants to select a longer radial length scale compared with the weakly nonlinear theory. Indeed, our early tests found that the diagnostics for the equilibrium sloshing state were sensitive to the box size, before converging for sufficiently wide boxes. This also allows for a finer sampling of quantized radial wavenumber, hence minimizing the possibility that certain mode couplings are missed. To maintain the same resolution as before, we also increase the number of cells so that $(N_x,N_y,N_z) = (1232,200,128)$. We start off from a Keplerian flow, with white-noise fluctuations in the density and velocity components at the level of 0.01, which act to seed the PI. We then run the experiment for $60T$, which tracks the initial resonant growth phase, the disruption due to the PI, and the subsequent equilibration toward a steady sloshing state moderated by turbulent hydrodynamic stresses. Throughout the analysis of our simulations, we will regularly take averages of quantities across our computational domain. We will denote spatial averages of a quantity $X$ as
\begin{equation}
    \langle{X\rangle}_{\Pi_i x_i} = \frac{\int X \,dD}{\int dD},
\end{equation}
whilst we denote density-weighted averages as
\begin{equation}
    \lbrace X \rbrace_{\Pi_i x_i} = \frac{\int \rho X \, dD}{{\int \rho \, dD}},
\end{equation}
where $dD = \Pi_i dx_i$ is the differential volume element for some combination of dimensions.

\subsection{Growth of Instability}
\label{subsec:instability_growth}

\begin{figure}
    \centering
    \includegraphics[width=\linewidth]{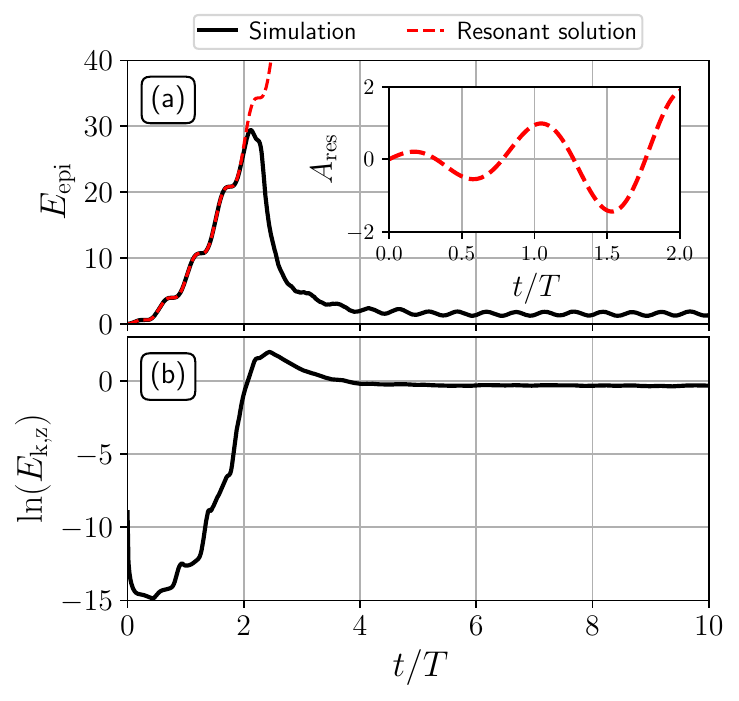}
    \caption{Growth of the PI for our $\psi_{\rm f} = 0.3$ hydrodynamic run. \textit{Panel (a)}: epicylic kinetic energy measured in the simulation (black line) and the result expected for the inviscid, laminar response (red dashed line). \textit{Inset panel}: the resonant growth of the sloshing motion $A_{\rm res}$ during early times. \textit{Panel (b)}: the logarithm of the box-integrated vertical kinetic energy.}
    \label{fig:figure_epicylic_growth}
\end{figure}

In Fig.~\ref{fig:figure_epicylic_growth} we track the evolution of the forced sloshing motion as a function of time. In particular, in panel (a), we are plotting the box-integrated epicylic kinetic energy, which is defined as
\begin{equation}
    E_{\rm epi} = L_x L_y L_z \Big\langle \frac{1}{2}\rho u_x^2+2\rho (u_y-u_{\rm K})^2 \Big\rangle_{xyz}.
\end{equation}
The black line shows the results measured directly from our simulation. Meanwhile, the red dashed line shows the analytical result arising from the inviscid resonant response, described in Section \ref{sec:problem_setup}. Indeed, according to this simplified laminar model, we can express the integrated epicylic energy in terms of the epicylic amplitude $C$, such that
\begin{equation}
    \label{eq:box_integrated_epi}
    E_{\rm epi} = \frac{L_x L_y L_z}{4}\rho_0 c^2 C,
\end{equation}
where $C = A^2+4B^2$ is computed from equations \eqref{eq:resonant_response_A}-\eqref{eq:resonant_response_B}. For example, the inset panel shows the growing response of the $x$ oscillation $A_{\rm res}$ over the first two orbital periods. At early times, we see very good agreement between these two curves as the epicylic motion is amplified in accordance with the laminar, resonant theory. However, at $t/T \sim 2$ we see the simulation result turnover as the epicylic amplitude is strongly quenched. The PI efficiently siphons energy from the coherent sloshing motion and redirects it toward growing inertial modes. This is clearly seen in panel (b), which shows the logarithm of the box-integrated vertical kinetic energy. This undergoes an approximately exponential growth phase, before saturating just after the peak in epicylic energy. At late times, both $E_{\rm epi}$ and $E_{{\rm k},z}$ settle down to a quasi-steady state. This represents the equilibrium sloshing state in which the coherent epicylic motions have come into balance with the nonlinear manifestation of the PI. 

\begin{figure}
    \centering
    \includegraphics[width=\linewidth]{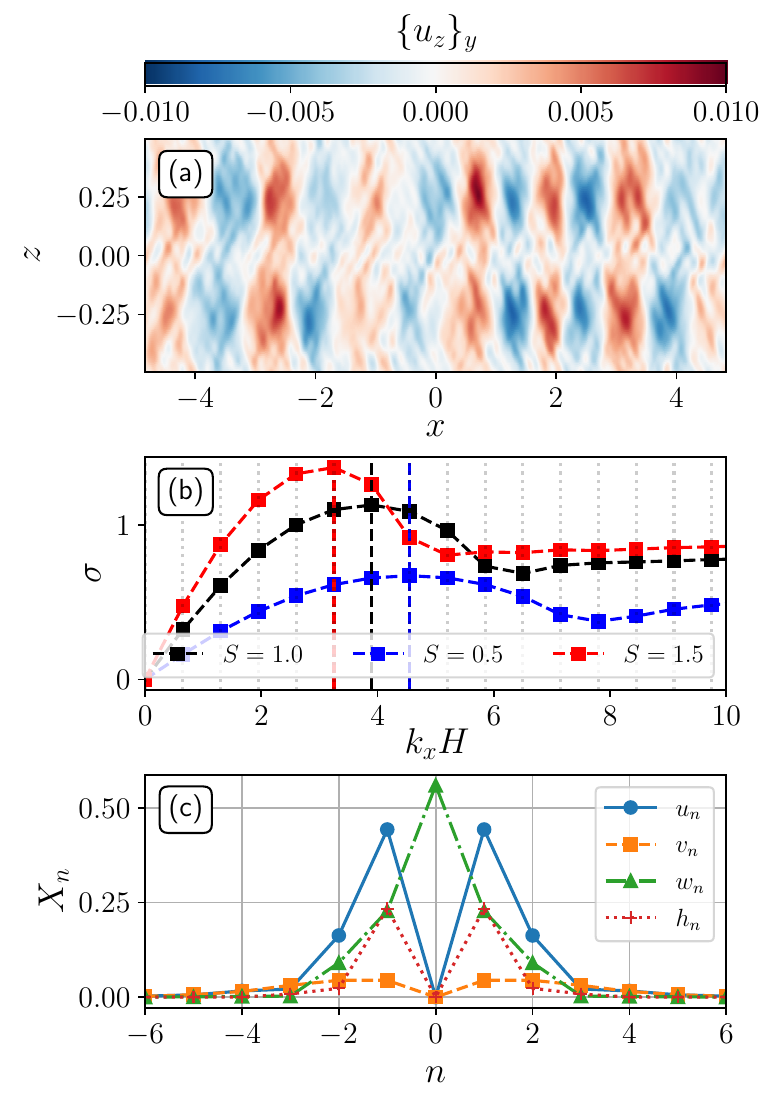}
    \caption{PI excited modal structure. \textit{Panel (a)}: meridional $x$--$z$ cross-section of $\lbrace u_z\rbrace_y$. \textit{Panel (b)}: maximal Floquet growth rates extracted as a function of the $k_x$ (at the allowed values quantized by the box) for different sloshing amplitudes $S = 0.5$ (blue), $1.0$ (black) and $1.5$ (red). \textit{Panel (c)}: the modal structure corresponding to the maximal growth rate for $S=1.0$. The $u_n$, $v_n$, $w_n$ and $h_n$ contributions are shown as blue circles, orange squares, green triangles, and red crosses, respectively.}
    \label{fig:figure_high_slosh_floquet}
\end{figure}

One should notice from the inset panel in Fig.~\ref{fig:figure_epicylic_growth}, that the radial flows become supersonic around $t \sim T$. This strong, and time-dependent, sloshing amplitude obviously departs from the realms of applicability for the three-mode coupling analysis discussed in Section \ref{subsec:WN_PI}. Therefore, we might expect the emergent PI to exhibit different properties. Indeed, the exponential growth rate of the vertical velocity in panel (b) is not as clean as that found in the controlled experiment shown in Fig.~\ref{fig:figure_PI_mode_growth}. Indeed, as $S$ becomes large, one should expect a transient modulation of the growing waves over a sloshing cycle. Moreover, the spatial structure of the growing perturbations is notably different. This is visualized in Fig.~\ref{fig:figure_high_slosh_floquet}, where panel (a) plots the density-weighted, $y$-averaged vertical velocity perturbations $\lbrace u_z\rbrace_y$ across the meridional extent of the box at $t/T = 1.5$. This shows a striking pattern of alternating vertical velocity channels that have a predominantly $n=0$ vertical structure, reminiscent of that found in Fig.~6 of \cite{FairbairnOgilvie_2023}. Hereafter, we will refer to these distinctive structures as `elevator' flows. In this snapshot, we also see that the pattern fits around 6 radial wavelengths into the box, which corresponds to a wavenumber of $k_x H = 3.9$. This is clearly a much longer wavelength compared with the smallest resonant wavenumber seen in the weakly nonlinear theory, exemplified in Fig.~\ref{fig:figure_floquet}. This once again underlines the importance of simulating a wide enough box that captures a number of these wavelengths. Upon testing different $L_x$, we find that the emergence of a preferred wavelength (which can change with time) is a robust result, supporting a physical origin over some scaling with the box width.

To better understand the emergence of the elevator flows, we can appeal to the use of the Floquet theory (see Section \ref{subsec:floquet}), which can still be applied for large values of $S$. Whilst this formalism strictly requires a time-periodic set of equations for which $S$ is constant, we find that it still maps effectively onto our full simulation results. In panel (b) of Fig.~\ref{fig:figure_high_slosh_floquet}, we plot the maximum growth rate computed from the Floquet theory as a function of $k_x$ using $|n|\leq 30$ modes. In particular, we extract these growth rates at the discrete values admitted by our periodic box. Since the sloshing amplitude increases with time during the resonant growth phase, we perform this analysis for three different values of $S = 0.5$, $1.0$, and $1.5$ (plotted by the blue, black, and red markers, respectively). We observe that the narrow resonant peaks, found in the three-mode coupling theory, are replaced by extended bands of instability \citep[also see][]{OgilvieLatter2013b, PaardekooperOgilvie_2019}. Moreover, we find peaks at much smaller wavenumbers, as indicated by the accordingly colored, dashed vertical lines. As the sloshing increases, the optimal mode moves toward longer wavelengths. This matches the behavior seen in the simulation, where we observe small-scale structure at early times, before longer wavelengths begin to dominate later. The optimal wavelength for $S=1.0$ occurs at $k_x H = 3.9$, which is in agreement with the snapshot in panel (a). Although the radial sloshing first reaches $A_{\rm res} = 1$ around $t/T \sim 1$, whilst panel (a) actually corresponds to $t/T=1.5$, it is reasonable to expect that there is some time delay for the preferred mode to grow to prominence. Of course, since the sloshing amplitude is continually changing, the preferred mode also evolves. Eventually, during the quasi-steady sloshing state, the vertical velocity settles into a structure with four radial wavelengths in the box. Finally, in panel (c), we plot the eigenmode structure for the $S=1.0$ optimal mode. The contribution from each of the vertical expansion coefficients is plotted as a function of $n$. The $w_0$ mode dominates the structure as the green, triangle markers peak at $n=0$. This pleasingly supports the development of the prominent $n=0$ elevator flows found in the simulations. These vertical flows enhance the communication between the horizontal, sloshing layers --  leading to hydrodynamic stresses which provide a rigidity, resisting the resonant forcing (see sections \ref{subsec:equilibrium_sloshing} and \ref{subsec:hydro_reynolds_stress}).

Similar upward and downward velocity channels have curiously emerged in a variety of other unstratified local simulation setups \citep[e.g.][]{DewberryEtAl_2020,TeedLatter_2021}, wherein they are vaguely attributed to generic forcing by inertial wave turbulence. These elevator modes might also be tentatively connected with the spectrum for oceanic internal waves, which is found to peak around the lowest frequency modes \citep{GarrettMunk_1979}.\footnote{We thank the referee for bringing this connection to our attention.} The physical picture is based on the preponderance for resonant triad interactions, which transfer energy from parent modes to children with lower frequencies \citep[][]{MullerEtAl_1986,ShavitEtAl_2025}. This weakly nonlinear wave turbulence can therefore establish an inverse cascade toward low-frequency modes. Whether this speculative mechanism helps to sustain the elevator modes in our context is beyond the scope of our present study, but ripe for future investigation.

\subsection{Equilibrium Sloshing}
\label{subsec:equilibrium_sloshing}

\begin{figure*}
    \centering
    \includegraphics[width=\linewidth]{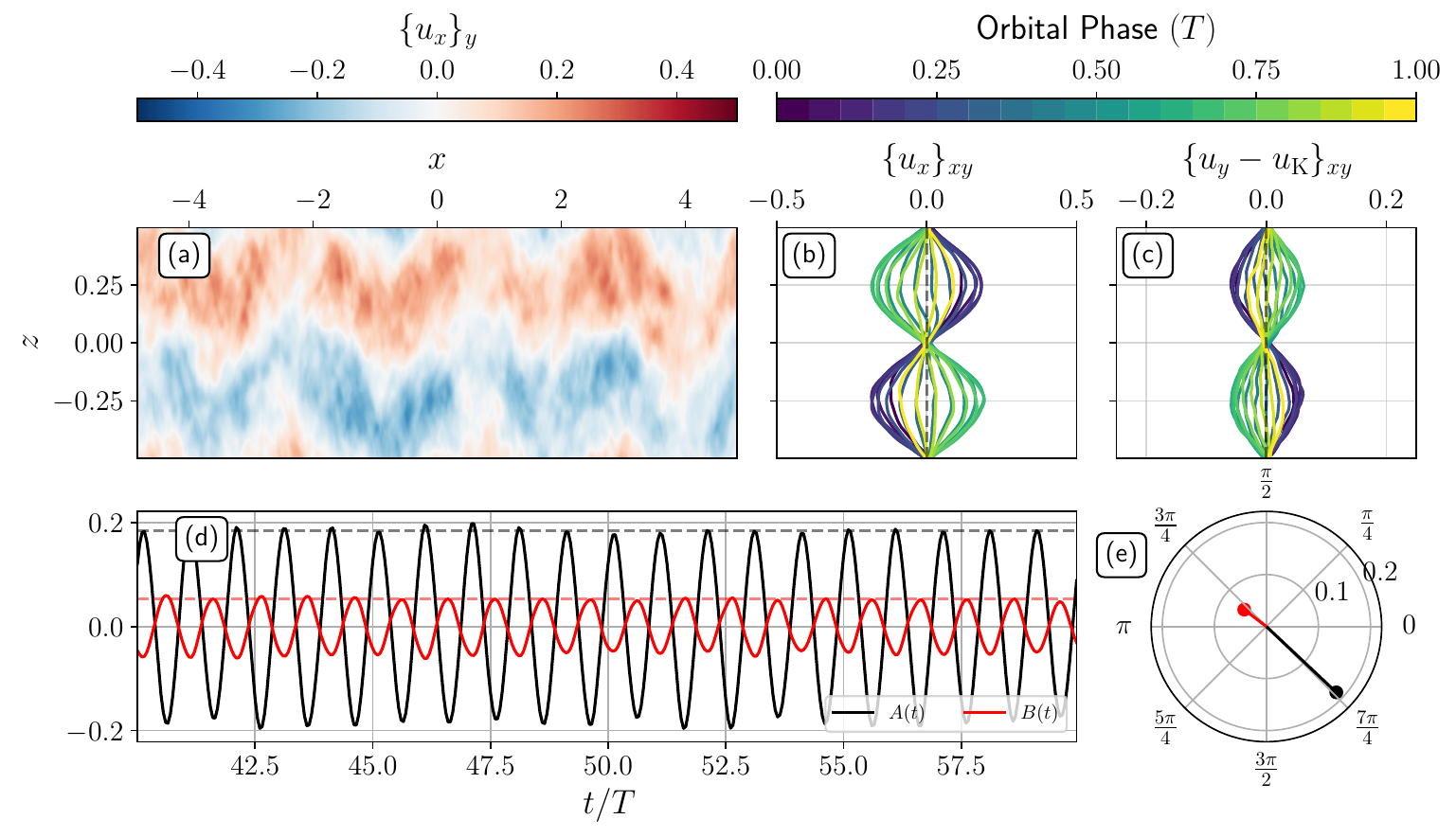}
    \caption{Sloshing equilibrium established for our $\psi_{\rm f} = 0.3$, hydrodynamic run. \textit{Panel (a)}: meridional cross section of $\lbrace u_x \rbrace_y$ at $t = 60T$. \textit{Panel (b)}: $\lbrace u_x \rbrace_{xy}$ profile as a function of $z$ for different oscillation phases over the orbital period $40 < t/T < 41$. \textit{Panel (c)}: likewise for $\lbrace u_y-u_{\rm K} \rbrace_{xy}$. \textit{Panel (d)}: best-fit sinusoidal profile for the radial sloshing $A(t)$ (black line) and azimuthal sloshing $B(t)$ (red line). \textit{Panel (e)}: polar plot of the Fourier amplitudes and phases of the extracted $A(t)$ (black) and $B(t)$ (red) oscillations, corresponding to the dynamical frequency.}
    \label{fig:figure_hydro_slosh_equilibrium}
\end{figure*}

After $t/T \sim 10$, the simulation broadly equilibrates into a regular sloshing back and forth, as indicated by the leveling off of the epicylic energy in panel (a) of Fig.~\ref{fig:figure_epicylic_growth}. We extend this run until $t/T = 60$ to ensure we have attained a quasi-steady state. The structure of this flow is visualized in Fig.~\ref{fig:figure_hydro_slosh_equilibrium}. In panel (a), we plot the density-weighted, $y$-averaged radial velocity $\lbrace u_x \rbrace_y$ at the final snapshot $t=60T$, across the meridional $x$--$z$ plane. Here, we clearly see the established sloshing flow, which is approximately odd about the midplane. However, we also recognize that this flow is distinct from the smooth, laminar picture presented in Section \ref{sec:problem_setup}. Indeed, the nonlinear manifestation of the PI distorts the interface between the sloshing layers as the strong vertical flows, discussed in Section \ref{subsec:instability_growth}, lead to a corrugation pattern. The magnitude of the sloshing flow is also notably reduced compared to the supersonic velocities attained during the resonant evolution, before the PI kicks in. In fact, we can obtain more quantitative information by performing a horizontal, density-weighted average of the non-Keplerian velocity field. We plot the $z$-profiles of $\lbrace u_x \rbrace_{xy} $ and $\lbrace u_y-u_K \rbrace_{xy}$ in panels (b) and (c), respectively. The colored lines show the sloshing profiles at different times during the orbital forcing phase (here plotted for the orbit $40<t/T<41$). These nicely exemplify the sinusoidal and oscillatory behavior of the averaged sloshing motions -- in keeping with a simplified, viscous, laminar model. For each snapshot, we use a least-squares procedure to fit the profiles with a sinusoidal function $\propto \sin(k_{\rm b} z)$ in order to extract the sloshing amplitudes $A(t)$ and $B(t)$. These are plotted as a function of time in panel (d), where the radial sloshing $A(t)$ and azimuthal sloshing $B(t)$ are shown as the black and red lines, respectively. These exhibit steady amplitudes and maintain a coherent, antiphased relationship. We perform a Fourier analysis on these time series and find the dominant frequency matches the orbital frequency, as expected. The Fourier component at this frequency measures the signal amplitude and phase as shown in the polar plot in panel (e). Writing $A(t) = \mathrm{Re}\lbrace|A|\exp[i(\arg(A)+\Omega_0 t)]\rbrace$ and likewise for $B(t)$, we find $|A| = 0.185$ and $\mathrm{arg}(A) = -0.76$, whilst $|B| = 0.054$ and $\mathrm{arg}(B) = 2.49$ -- yielding an amplitude ratio of $3.43$ and a phase difference of $1.03\pi$.

We can test how well this behavior is captured by the laminar, viscous model presented in Section \ref{sec:problem_setup}. In particular, this model predicts an equilibrium described by equations \eqref{eq:viscous_response_A}-\eqref{eq:viscous_response_B}, which encapsulate the amplitude and phase relationship of the $x$ and $y$ sloshing motions. Since we have two free parameters, $\nu_{xz}$ and $\nu_{yz}$, we will try to optimize these by minimizing the cost function
\begin{equation}
    \mathcal{C}(\nu_{xz},\nu_{yz}) = \left(\frac{|A|-|A|_{\rm vis}}{|A|}\right)^2+\left(\frac{|B|-|B|_{\rm vis}}{|B|}\right)^2.
\end{equation}
\begin{figure}
    \centering
    \includegraphics[width=\linewidth]{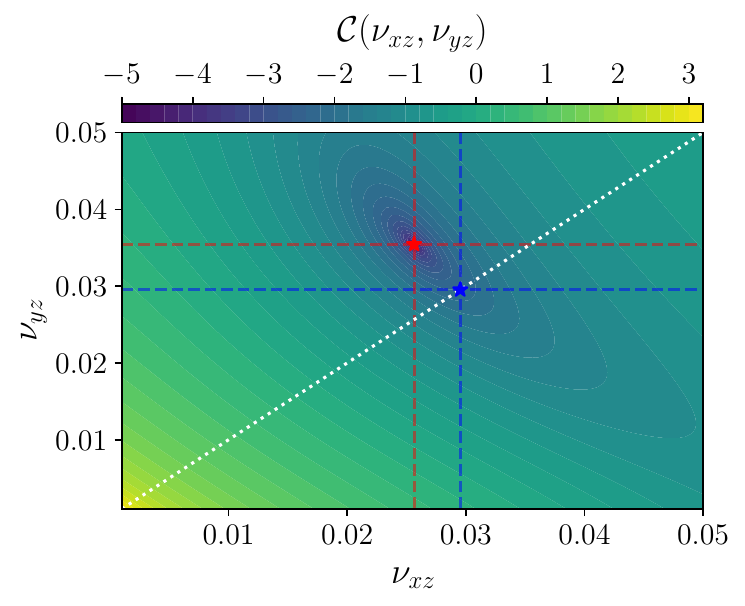}
    \caption{Optimization of the viscosity cost function $\mathcal{C}(\nu_{xz},\nu_{yz})$ to fit $|A|$ and $|B|$. The absolute minimum is located at the red star where $(\nu_{xz},\nu_{yz}) = (0.0256,0.0354)$. When constrained to an isotropic viscosity, along the dotted white line, the minimum is located at the blue star for which $\nu_{xy} = \nu_{yz} = 0.0295$.}
    \label{fig:figure_viscous_fit_contour}
\end{figure}
The results are shown in Fig.~\ref{fig:figure_viscous_fit_contour}, wherein the contour plot visualizes level sets of the cost function. This is minimized at the location of the red star for which $(\nu_{xz},\nu_{yz}) = (0.0256,0.0354)$. This yields $|A|_{\rm vis} = 0.185$ and $|B|_{\rm vis} = 0.054$ in accordance with the amplitudes extracted from the simulation. However, the resulting model phase difference $\textrm{arg}(B_{\rm vis})-\textrm{arg}(A_{\rm vis}) = 0.802\pi$ is clearly different. This disparity is indicative of the shortcomings of a viscous, shearing prescription in capturing the full details of a turbulent, stress relationship. Even more simply, if we restrict our attention to isotropic viscosities along the white dashed line, we find that the cost function is minimized for $\nu_{xy} = \nu_{yz} = 0.0295$ at the position of the blue star.\footnote{Notice that the value of $\nu$ extracted here is too large to apply the simpler formula given by equation \eqref{eq:simple_alpha}.} This simplified isotropic model then predicts $|A|_{\rm vis} = 0.171$ and $|B|_{\rm vis} = 0.056$, with $\textrm{arg}(B_{\rm vis})-\textrm{arg}(A_{\rm vis}) = 0.774\pi$, which is a marginally worse fit to the simulation results, compared with the anisotropic model considered above. This is unsurprising since the isotropic assumption reduces the available degrees of freedom for fitting the data. Moreover, as we will see in the following section, these purely viscous prescriptions will be supplanted by more flexible stress-strain models, which perform even better at fitting the data.

\subsection{Reynolds Stresses}
\label{subsec:hydro_reynolds_stress}

In fact, we can be more quantitative about the nature of the turbulent stresses established in our simulation. Upon horizontally averaging the inviscid, hydrodynamic momentum equation \eqref{eq:momentum}, we arrive at 
\begin{eqnarray}
    \label{eq:mom_x_avg}
    && \frac{\partial \langle \rho v_x\rangle_{xy}}{\partial t}-2\Omega_0\langle \rho v_y\rangle_{xy} = \langle f_{\rm w}\rangle_{xy}-\frac{\partial}{\partial z}\langle \rho v_x v_z\rangle_{xy}  , \\
    \label{eq:mom_y_avg}
    && \frac{\partial \langle \rho v_y\rangle_{xy}}{\partial t}+\frac{1}{2}\Omega_0\langle \rho v_x\rangle_{xy} =-\frac{\partial}{\partial z}\langle \rho v_y v_z\rangle_{xy},
\end{eqnarray}
akin to equations (8)-(9) of \cite{TorkelssonEtAl_2000} but with the additional forcing term. Here, we work with the velocity components $\mathbf{v} = \mathbf{u}-\mathbf{u}_{\rm K}$, relative to the underlying Keplerian flow. The horizontal averaging extracts the coherent sloshing motion, which we will denote $\mathbf{v}_{\rm w} \equiv \langle\rho \mathbf{v}\rangle_{xy}/\bar{\rho} = \lbrace \mathbf{v}\rbrace_{xy}$, where the density average $\bar{\rho} = \langle \rho\rangle_{xy}$. Note that we can identify the in-plane, averaged components with the sloshing amplitudes $A$ and $B$ as defined in equations \eqref{eq:laminar_ansatz_A}-\eqref{eq:laminar_ansatz_B}. Therefore, we decompose the non-Keplerian velocity into a sloshing part and a turbulent residual $\mathbf{u}^\prime$ such that $\mathbf{v} = \mathbf{v}_{\rm w}+\mathbf{u}^\prime$, where $\lbrace\mathbf{u}^\prime\rbrace_{xy} = 0$. Propagating this splitting through equations \eqref{eq:mom_x_avg}-\eqref{eq:mom_y_avg}, we find
\begin{eqnarray}
    \label{eq:mom_x_rey_avg}
    && \frac{\partial \bar{\rho}{v}_{x,{\rm w}}}{\partial t}-2\Omega_0\bar{\rho}v_{y,{\rm w}} = \langle f_{\rm w}\rangle-\frac{\partial \mathcal{R}_{xz}}{\partial z}  , \\
    \label{eq:mom_y_rey_avg}
    && \frac{\partial \bar{\rho}v_{y,{\rm w}}}{\partial t}+\frac{1}{2}\Omega_0\bar{\rho}v_{x,{\rm w}} =-\frac{\partial \mathcal{R}_{yz}}{\partial z},
\end{eqnarray}
where the turbulent Reynolds stresses are given by
\begin{equation}
    \mathcal{R}_{ij} \equiv \langle \rho u_i^\prime u_j^\prime \rangle = \bar{\rho}\lbrace u_i^\prime u_j^\prime\rbrace.
\end{equation}
This formal averaging procedure demonstrates that the fully turbulent equations can be effectively reduced to the laminar problem described by equations \eqref{eq:Adot_eqn}-\eqref{eq:Bdot_eqn}, where the viscous terms are supplanted by the Reynolds stresses. 

\begin{figure}
    \centering
    \includegraphics[width=\linewidth]{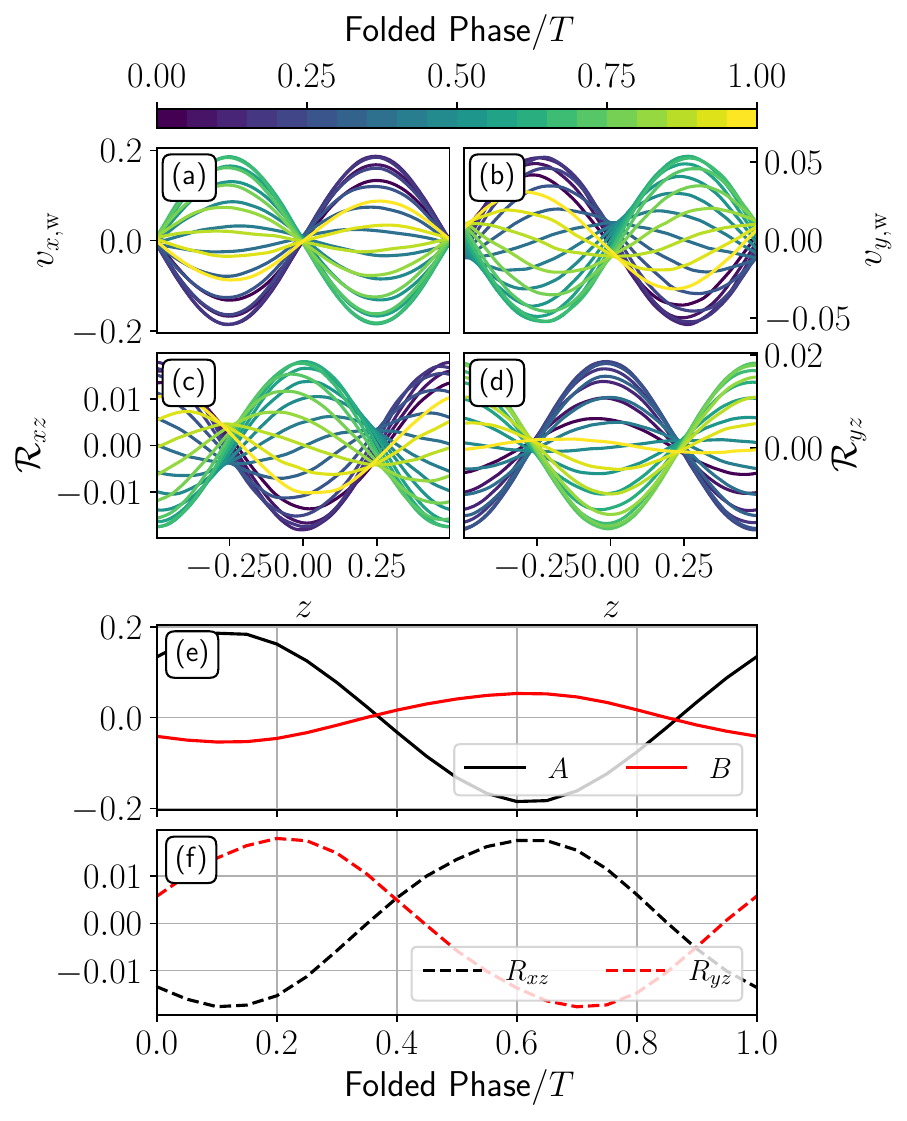}
    \caption{The rate of strain and stress relationship for our fiducial hydrodynamic run. The upper panels show the phase-folded, vertical profiles for: (a) $v_{x,{\rm w}}$, (b) $v_{y,{\rm w}}$, (c) $\mathcal{R}_{xz}$, and (d) $\mathcal{R}_{yz}$. In panel (e), we extract the best-fitting, sloshing amplitudes $A(t)$ (black) and $B(t)$ (red) across the orbital phase. In panel (f), we similarly extract the best-fitting stress amplitudes, $R_{xz}(t)$ (black-dashed) and $R_{yz}(t)$ (red-dashed).}
    \label{fig:figure_hydro_reynolds}
\end{figure}

In Section \ref{subsec:equilibrium_sloshing}, we examined the simplest viscous closure for setting the equilibrium sloshing behavior. Now we are able to directly interrogate the relationship between these averaged flows and the hydrodynamic stress to educate an improved closure model for the turbulence. In Fig.~\ref{fig:figure_hydro_reynolds}, we compare the sloshing and stress response. In the upper panels (a) and (b), we plot $v_{x,{\rm w}}$ and $v_{y,{\rm w}}$ as a function of $z$ over an orbital forcing period, where we have phase-folded in the profiles from all snapshots during the interval $40 < t/T <60$. Similarly, we perform this procedure with the Reynolds stresses, $\mathcal{R}_{xy}$ and $\mathcal{R}_{yz}$, which are shown in panels (c) and (d), respectively. The sinusoidal profile for the bulk sloshing motion recapitulates the findings seen in panels (b) and (c) of Fig.~\ref{fig:figure_hydro_slosh_equilibrium}. Meanwhile, we see that the Reynolds stresses develop an obvious cosine structure, with extrema collocated with the maximally shearing layers. Indeed, the vertical derivative of this Reynolds stress ensures that the overall spatial dependence cancels in equations \eqref{eq:mom_x_rey_avg}-\eqref{eq:mom_y_rey_avg}. At each of the orbital phases, we can fit a sine function to the sloshing motion to extract $A(t)$ and $B(t)$, akin to panel (d) in Fig.~\ref{fig:figure_hydro_slosh_equilibrium}, but now for our phase-folded data. These are shown in panel (e) of Fig.~\ref{fig:figure_hydro_reynolds}. Similarly, we fit the Reynolds stresses according to $\mathcal{R}_{ij}= R_{ij}(t)\cos(k_{\rm b} z)$, which are shown in panel (f). Fourier analyzing these curves, we extract the associated amplitudes and phases of the dominant $\Omega_0$ frequency component. In keeping with our previous results, $|A| = 0.185$ and $\arg(A) = -0.76$ whilst $|B| = 0.054$ and $\arg(B) = 2.49$.\footnote{By linearity, the Fourier series for a signal over many cycles is equal to the phase-folded, average signal over a single cycle.} Meanwhile, we find that $|R_{xz}| = 0.0178$  and $\arg(R_{xz}) = 2.48$, whilst  $|R_{yz}| = 0.0177$  and $\arg(R_{yz}) = -1.26$. For the previously assumed, purely viscous model, one typically equates $-\mathcal{R}_{ij} = \bar{\rho}\nu_{ij}\partial v_{i,{\rm w}}/\partial z$. Inserting the ansatz for the stress-slosh spatial structure yields
\begin{equation}
    \label{eq:rey_stress-slosh}
    R_{xz} = -\bar{\rho} \nu_{xz} c k_{\rm b} A \quad \textrm{and} \quad R_{yz} = -\bar{\rho} \nu_{yz} c k_{\rm b} B.
\end{equation}
This model suggests that the Reynolds stresses should be $\pi$ out of phase with the sloshing flows. In fact, we see that $\arg(R_{xz})-\arg(A) = 1.03\pi$ and $\arg(R_{yz})-\arg(B) = -1.19\pi$. Although these are relatively consistent with the viscous picture, there are clear discrepancies. This reemphasizes the results of Section \ref{subsec:equilibrium_sloshing}, wherein the best viscous fit for the equilibrium sloshing could readily capture the sloshing amplitudes, but not their phase relationship. Therefore, we are encouraged to relax the purely viscous prescription and instead appeal to a more flexible closure model for the turbulence. Indeed, by assuming that $\nu$ is real, we are implicitly stating that the stress responds instantaneously to the rate of strain. More realistically, the turbulent transport depends on the history of the shearing motion -- in other words, there is some relaxation or delay timescale between the shear and the stress response. Such time delays are also reminiscent of local simulations of the MRI, wherein small lags are found between energy injection rates, via the boundary stresses, and the thermalization rate, as channel flows take some finite time to subsequently decay via parasitic instabilities \citep[see][]{SanoInutsuka_2001,GardinerStone_2005a}. This suggests the use of a \textit{viscoelastic} model for which the stress and strain rate can have an arbitrary phase relationship \citep[e.g.][]{OgilvieProctor_2003,OgilvieLesur_2012}. We therefore promote $\nu$ to being a complex quantity. Solving equation \eqref{eq:rey_stress-slosh} for the viscoelastic components gives $\nu_{xz} = 0.015+0.001i = 0.015e^{0.09i}$ and $\nu_{yz} = 0.043-0.030i = 0.053e^{-0.60i}$. With this complex viscosity in hand, we can also generalize the predictions of the steady-state, laminar sloshing given by equations \eqref{eq:viscous_response_A}-\eqref{eq:viscous_response_B}. Inserting our new results for $\nu_{xz}$ and $\nu_{yz}$ we find that $|A|_{\rm vis} = 0.186$ and $|B|_{\rm vis} = 0.054$ with a phase difference of $\arg(B_{\rm vis})-\arg(A_{\rm vis}) = 1.03\pi$, almost in perfect agreement with the values extracted from the simulation. This should be expected given the added flexibility of the viscoelastic model, where each coefficient $\nu_{iz}$ provides two degrees of freedom, which can then fit two amplitudes and two phases.

%% file: 4_mhd_slosh.tex
\section{MHD results}
\label{sec:mhd_slosh}

Having investigated the steady sloshing response for the hydrodynamic case, we now wish to extend our setup to include magnetic fields. This will facilitate a complicated interplay between the MRI and PI, which jointly determines the final sloshing state equilibrium. In Section \ref{subsection:base_mri}, we will characterize the base MRI state before the addition of sloshing flows. Then, in Section \ref{subsection:PI_MRI_growth}, we will examine how the growth of the PI proceeds in the presence of MRI turbulence. Next, in Section \ref{subsection:MHD_vertical_stresses}, we will investigate the resulting equilibrium sloshing state and quantify the vertical magnetic and hydrodynamic stresses at play. Finally, in Section \ref{subsec:forcing_amplitude}, we will interrogate the dependence of our results on the forcing amplitude, comparing both the hydrodynamic and magnetized regimes.

\subsection{Characterizing the Base MRI}
\label{subsection:base_mri}

Before implementing the sloshing forcing, we first need to develop a steady MRI base state in our shearing box. To this end, we will extend our use of \texttt{AthenaK} to solve equations \eqref{eq:continuity}-\eqref{eq:induction} in their full generality, reintroducing the magnetic field. We initialize the magnetic field with a zero-net-flux profile according to 
\begin{equation}
    B_x = B_y = 0 \quad \textrm{and} \quad  B_z = B_0 \sin(2\pi x/L_x),
\end{equation}
where $B_0 = \sqrt{2p_0/\beta_{\rm m}^2}$ and the plasma-beta parameter is chosen to be $\beta_{\rm m} = 400$.\footnote{Future experimentation should investigate the effect of different MRI field geometries.} Furthermore, since it is known that the saturation of the MRI depends on the viscous and ohmic diffusivity values, we explicitly set these to be $\nu = 3.2\times10^{-4}$ and $\eta = 8.0\times10^{-5}$, giving a magnetic Prandtl number $\textrm{Pd} \equiv \nu/\eta = 4$, in keeping with the fiducial setup examined by \cite{FromangEtAl_2007}. As per our hydrodynamic setup, we perform these simulations on a grid with $(L_x,L_y,L_z) = (9.6672, \pi, 1)$ and $(N_x, N_y, N_z) = (1232, 200, 128)$. We seed the instability with white-noise, relative pressure perturbations of amplitude $0.01$, and then run the simulation for $100T$ -- tracking the initial growth, nonlinear saturation, and final quasi-steady state for the MRI.

\begin{figure}
    \centering
    \includegraphics[width=\linewidth]{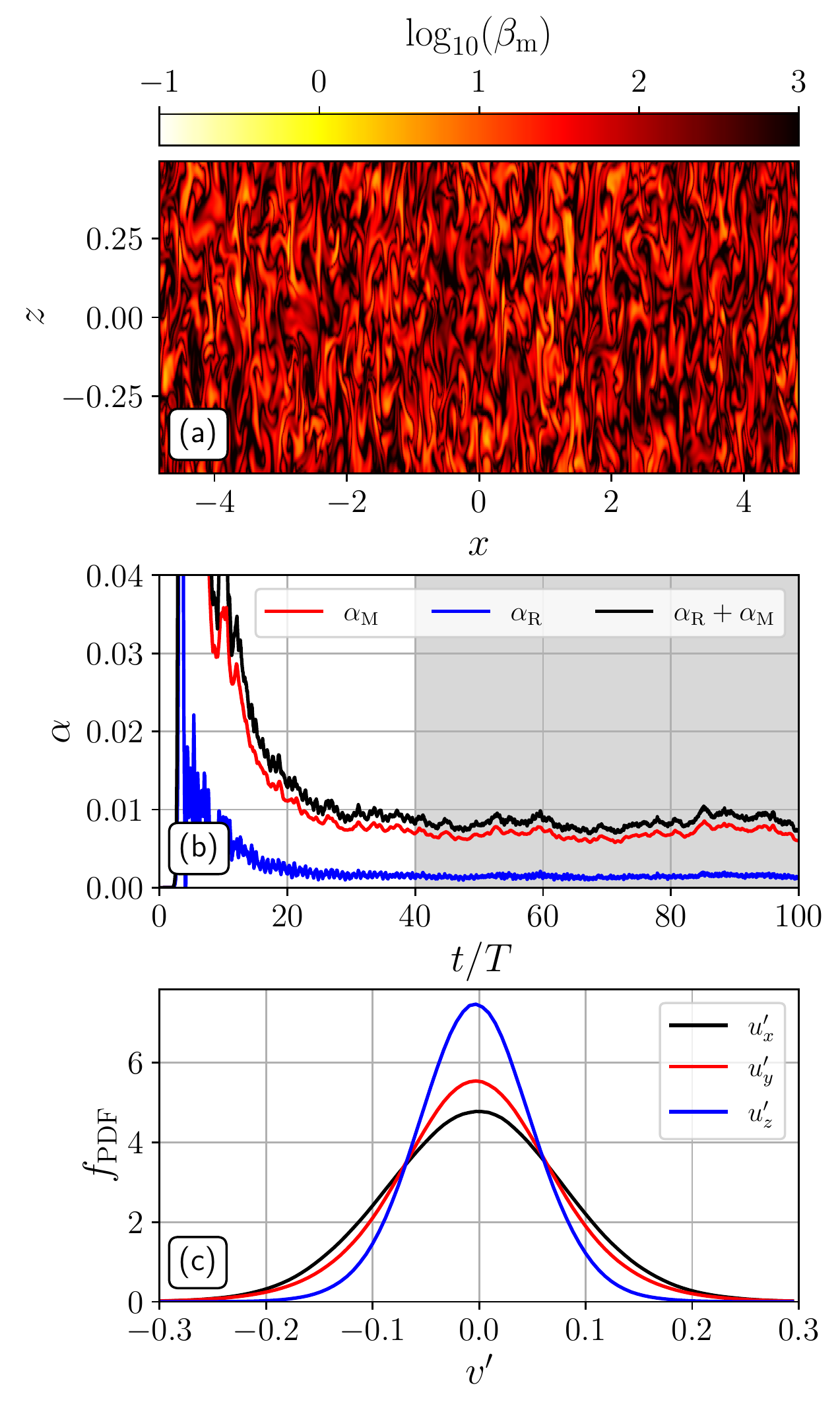}
    \caption{Characterization of our zero-net-flux, base MRI state. \textit{Panel (a)}: plots $\log_{10}(\beta_{\rm m})$ through the meridional slice $y=0$, showing field amplification and fine turbulent structure. \textit{Panel (b)}: growth and saturation of the horizontal Maxwell stresses $\alpha_{\rm M}$ (red), Reynolds stresses $\alpha_{\rm R}$ (blue) and net stress (black), all normalized by the box-averaged pressure. \textit{Panel (c)}: the density-weighted probability density function ($f_{\rm PDF}$) for the velocity fluctuations $u_{x}^\prime$ (black), $u_{y}^\prime$ (red) and $u_{z}^\prime$ (blue), averaged over simulation snapshots between $40 < t/T < 100$ in intervals of 10.}
    \label{fig:figure_MRI_state}
\end{figure}

Representative summary diagnostics, characterizing the base MRI state, are shown in Fig.~\ref{fig:figure_MRI_state}. In panel (a), we plot $\log_{10}(\beta_{\rm m})$ through the $y=0$ slice at $t = 100T$, once the MRI has saturated. Here, we clearly observe the presence of turbulent structure over a range of scales. Furthermore, amplification of the magnetic field leads to eddies in which $\beta_{\rm m}$ is of order unity. The strength of the MRI is typically quantified in terms of the nondimensionalized, horizontal accretion stresses it induces. These consist of the hydrodynamic Reynolds stresses
\begin{equation}
    \alpha_{\rm R} = \frac{\langle \rho u_x^\prime u_y^\prime \rangle_{xyz}}{q \langle \rho c^2 \rangle_{xyz}} = \frac{\lbrace u_x^\prime u_y^\prime \rbrace_{xyz}}{q c^2},
\end{equation}
and the magnetic Maxwell stresses
\begin{equation}
    \alpha_{\rm M} = \frac{\langle B_x B_y \rangle_{xyz}}{q \langle \rho c^2 \rangle_{xyz}}.
\end{equation}
Here, we have normalized the stresses against the box-averaged pressure $\langle \rho c^2 \rangle = \rho_0 c^2 = p_0 =1$ and also the shear coefficient $q$, so that $\alpha$ can be identified with the Shakura-Sunyaev viscosity parameter acting on the horizontal shear flow. These average stresses are plotted as a function of time in panel (b), where $\alpha_{\rm R}$ (red line) and $\alpha_{\rm M}$ (blue line) are combined into the net stress (black line). Here we see a steep initial rise phase as the MRI grows over tens of dynamical orbits. As the instability saturates, the stresses turn over before leveling off beyond $t = 40 T$. The stresses are averaged over the gray shaded interval between $40 < t/T < 100$, yielding typical values for $\alpha_{\rm M} = 0.0069$ and $\alpha_{\rm R} = 0.0014$. As per usual, the Maxwell stresses dominate over the hydrodynamic effects. In the final panel (c), we plot the density-weighted, normalized histogram ($f_{\rm PDF}$) for the velocity fluctuations, averaged over snapshots at $t/T \in [40,100]$ in intervals of 10. Here we see that the horizontal fluctuations tend to have a wider variance, with characteristic standard deviations $\sigma_x  = 0.085$ and $\sigma_y = 0.079$, versus $\sigma_z = 0.056$ for the vertical velocity. For a strong sloshing signal to visibly emerge in the forced simulations, it should manifest on velocities larger than these. However, even small sloshing amplitudes, disguised by the turbulence, can be extracted in an average sense.

\subsection{Interaction of MRI and PI}
\label{subsection:PI_MRI_growth}

\begin{figure*}
    \centering
    \includegraphics[width=\linewidth]{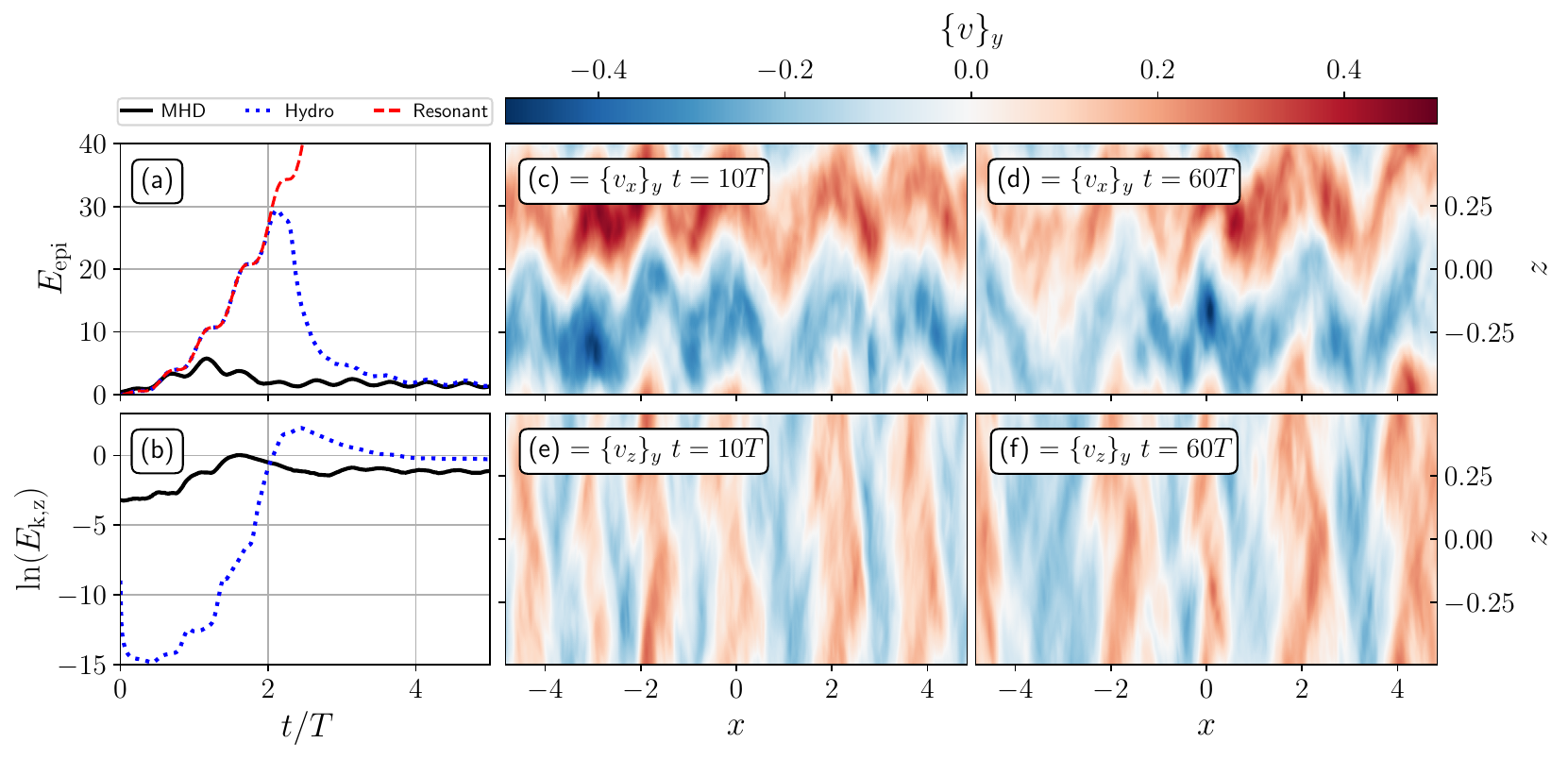}
    \caption{The emergence of the PI atop the MRI for our magnetized $\psi_{\rm f} = 0.3$ run. \textit{Panel (a)}: The growth of $E_{\rm epi}$ is shown for the MHD run (black solid) compared with the hydrodynamic run (blue dotted) from Section \ref{sec:hydro_slosh}. Once again, the laminar, resonant prediction is shown as the red dashed line. \textit{Panel (b)}: the rise in $\ln(E_{{\rm k},z})$ for the MHD (black solid) versus hydrodynamic (blue dotted) runs. The $y$-averaged radial sloshing flows $\lbrace v_x \rbrace_y$ at (c) $t = 10T$ and (d) $t = 60T$. The $y$-averaged vertical flows $\lbrace v_z \rbrace_y$ at (e) $t = 10T$ and (f) $t = 60T$.}
    \label{fig:figure_mri_epicylic_growth}
\end{figure*}

Having developed the base state MRI, we restart our simulation from the final snapshot output in Section \ref{subsection:base_mri}. As per our hydrodynamic sloshing run, we now switch on the sloshing forcing with a fiducial value of $\psi_{\rm f} = 0.3$. We run the simulation for an additional $60T$, which tracks the growth of the PI on top of the MRI, and the eventual saturation into a quasi-steady sloshing state. The clear development of the PI is once again seen in Fig.~\ref{fig:figure_mri_epicylic_growth}.
Panels (a) and (b) show the growth of $E_{\rm epi}$ and $\ln(E_{{\rm k},z})$, similarly to Fig.~\ref{fig:figure_epicylic_growth}. Now, the new MHD simulation results are shown in black, whilst the previous hydrodynamic results are shown as the blue dotted line. The red dashed line shows the laminar resonant amplification of the sloshing motion, as before. Here we see that the initial epicylic growth is less pronounced. The MRI turbulence naturally resists the resonant growth and also seeds stronger fluctuations for the PI to latch onto. Together, this reduces the initial growth of the epicylic motion, and the PI takes over at earlier times -- as indicated by the early rise phase of the vertical kinetic energy. Nonetheless, the maximum radial sloshing motions still reach an appreciable fraction of the sound speed, and the ensuing PI has similar characteristics to the hydrodynamic run. Indeed, we see at late times that the epicylic energy drops down to a quasi-steady state, where the black and blue dotted lines apparently overlap. This already hints that the final sloshing state is fairly similar in both the magnetized versus hydrodynamic setups. This is also supported by the velocity maps, which are plotted for snapshots at $t=10T$ ((c) and (e)) and $t=60T$ ((d) and (f)). In panels (c) and (d), we plot $\lbrace v_x \rbrace_y$, which shows the clear presence of a sloshing flow that is broadly odd about the midplane, albeit with a distorted interface. This corrugation is once again driven by the emergence of a series of strong, counter-propagating vertical flows as visualized by $\lbrace v_z \rbrace_y$ in panels (e) and (f). The characteristic radial wavenumber in panel (d) is the same as that seen in Fig.~\ref{fig:figure_high_slosh_floquet}, which is indicative of a similar, strong sloshing PI mechanism at play. However, we note that by the later snapshot (f), the radial wavenumber has decreased, indicating that the established elevator modes can still slowly evolve. Indeed, examining the intermediate times shows the merging of two channels, and the preference for this longer wavelength -- perhaps aided by the MRI turbulence which diffuses and damps higher radial wavenumbers. Combining a model for the damping, facilitated by the MRI, with the PI Floquet growth rates for sufficiently subsonic sloshing amplitudes does indeed suggest that there is a preference for a longer wavelength to emerge at later times. This model will be discussed in more detail in Section \ref{subsec:forcing_amplitude}.

\begin{figure}
    \centering
    \includegraphics[width=\linewidth]{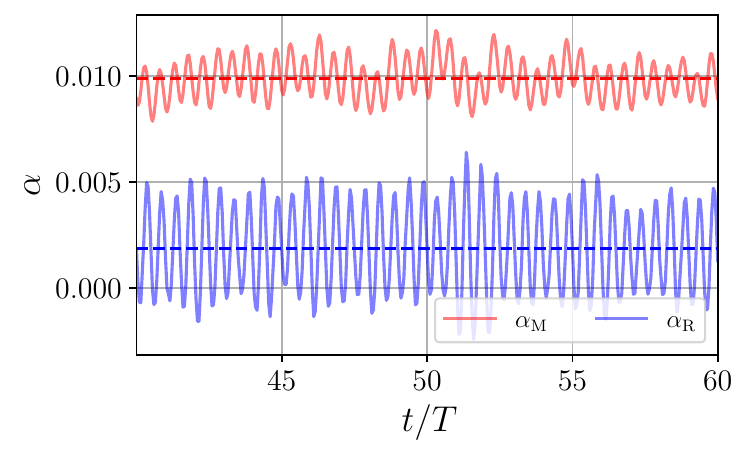}
    \caption{Horizontal stresses in the equilibrium sloshing state, measured between $40 < t/T <60$. $\alpha_{\rm M}$ (red line) oscillates about the averaged value of $0.0099$ (dashed red). $\alpha_{\rm R}$ (blue line) oscillates about the averaged value of $0.0019$ (dashed blue).}
    \label{fig:figure_slosh_alpha_xy}
\end{figure}

The sloshing state also modifies the nature of the horizontal stresses, compared with the standard MRI discussed in Section \ref{subsection:base_mri}. Note that when computing $\alpha_{\rm R} = \langle \rho u_x^\prime u_y^\prime\rangle_{xyz}/(q \langle \rho c^2 \rangle_{xyz})$, the velocity fluctuations are now measured relative to the average sloshing flow. The results are shown in Fig.~\ref{fig:figure_slosh_alpha_xy} over the time interval $40<t/T<60$, after the sloshing flow has equilibrated. The Maxwell stresses $\alpha_{\rm M}$ (red) once again dominate over the Reynolds stresses $\alpha_{\rm R}$ (blue), with time-averaged values of $0.0099$ and $0.0019$, respectively. This corresponds to an amplification of $\sim 1.4$ compared with the base state. It should also be noted that the horizontal stresses exhibit a clear oscillation over the dynamical timescale, indicating that the horizontal fluctuations are also correlated with the forcing frequency.

\subsection{MHD Vertical Stresses}
\label{subsection:MHD_vertical_stresses}

We will now quantify the MHD equilibrium sloshing state in detail, akin to the analysis performed in Section \ref{subsec:hydro_reynolds_stress}. In particular, we will examine the balance between the sloshing flows and the vertical stresses, which now comprise both Reynolds and Maxwell terms. The Maxwell stresses are denoted
\begin{equation}
    \label{eq:M_ij}
    \mathcal{M}_{ij} = -\langle B_i B_j\rangle_{xy} , 
\end{equation}
and arise from the horizontal averaging of equation \eqref{eq:momentum}. In this way, we extend our horizontally averaged equations for the sloshing motion by appending $-\partial \mathcal{M}_{xz}/\partial z$ and $-\partial \mathcal{M}_{yz}/\partial z$ to equations \eqref{eq:mom_x_rey_avg}-\eqref{eq:mom_y_rey_avg}, respectively. 

\begin{figure*}
    \centering
    \includegraphics[width=\linewidth]{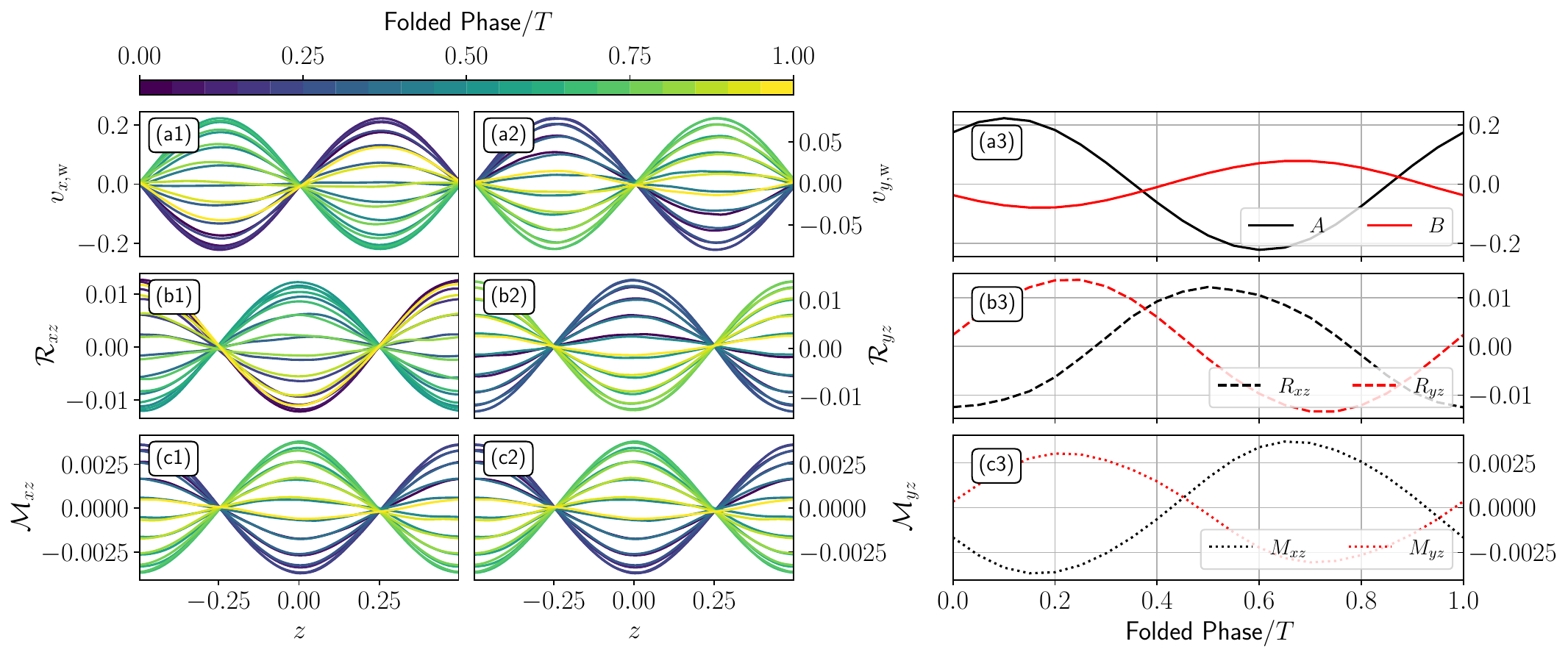}
    \caption{Rate of strain and stress relationship for our fiducial MHD run (akin to Fig.~\ref{fig:figure_hydro_reynolds}). Columns (1) and (2) respectively show the $x$ and $y$ components of the phase-folded, vertical profiles for the (a) sloshing motions, (b) Reynolds stresses, and (c) Maxwell stresses, at different orbital phases indicated by the line color. Column (3) then extracts the corresponding amplitudes of these sinusoidal profiles as a function of the temporal phase.}
    \label{fig:figure_mhd_reynolds_maxwell}
\end{figure*}

In Fig.~\ref{fig:figure_mhd_reynolds_maxwell}, we compare the amplitude and phase relationship between the equilibrium sloshing motion $\mathcal{R}$ and $\mathcal{M}$. As per the analysis presented in Fig.~\ref{fig:figure_hydro_reynolds}, we phase fold our data modulo the dynamical period, over the interval $40<t/T<60$, to boost our sloshing signal. In panels (a1) and (a2), we are plotting the horizontally averaged sloshing motion $v_{x,{\rm w}}$ and $v_{y,{\rm w}}$, as a function of $z$, across a number of orbital phases. Similarly, we extract the phase-folded $z$-profiles for $\mathcal{R}_{xz}$, $\mathcal{R}_{yz}$, $\mathcal{M}_{xz}$, and $\mathcal{M}_{yz}$ in panels (b1), (b2), (c1), and (c2). In keeping with the hydrodynamic results, we cleanly extract oscillating sinusoidal profiles for the sloshing motion and $\cos(k_{\rm b} z)$ structures for the stresses. At each phase, we find the best fit to these profiles and thereby extract $A(t)$, $B(t)$ (panel (a3)), $R_{ij}$ (panel (b3)), and $M_{ij}$ (panel (c3)), where $\mathcal{M}_{ij} = M_{ij}(t) \cos{(k_{\rm b}z)}$.

By Fourier analyzing these curves, we isolate the amplitude and phase for the dominant orbital frequency component. We find that $|A| = 0.221$ and $\arg(A) = -0.66$, whilst $|B| = 0.079$ and $\arg(B) = 2.06$. This shows that the sloshing amplitude is slightly larger in this case, compared with the previous hydrodynamic run. Performing the same analysis on the stresses, we find $|R_{xz}| = 0.0125$ and $\arg(R_{xz})=2.98$, whilst $|R_{yz}| = 0.0134$ and $\arg(R_{yz})=-1.40$. Finally, $|M_{xz}| = 0.0037$ and $\arg(M_{xz})=2.04$, whilst $|M_{yz}| = 0.0031$ and $\arg(M_{yz})=-1.44$. Clearly, the Reynolds stresses dominate the vertical transport, owing to the dominance of the PI in establishing strong elevator flows that resist the sloshing motions. 

Once again, we can find the viscoelastic linear relationship between the velocity shear and the stresses, such that $-\mathcal{R}_{ij} = \bar{\rho}\nu_{{\rm R},ij}\partial v_{i,\rm{w}}/\partial z$ and  $-\mathcal{M}_{ij} = \bar{\rho}\nu_{{\rm M},ij}\partial v_{i,\rm{w}}/\partial z$. In addition to equation \eqref{eq:rey_stress-slosh}, we have  an analogous expression involving the Maxwell viscoelastic coefficients
\begin{equation}
    \label{eq:max_stress-slosh}
    M_{xz} = -\bar{\rho} \nu_{{\rm M},xz} c k_{\rm b} A \quad \textrm{and} \quad M_{yz} = -\bar{\rho} \nu_{{\rm M},yz} c k_{\rm b} B.
\end{equation}
We find that $\nu_{{\rm R},xz} = 0.0079+0.0043i = 0.0090e^{0.50i}$ and $\nu_{{\rm R},yz} = 0.0261-0.0087i = 0.027e^{-0.32i}$. Meanwhile, $\nu_{{\rm M},xz} = 0.0024-0.0011i = 0.0027e^{-0.44i}$ and $\nu_{{\rm M},yz} = 0.0058-0.0021i = 0.0061e^{-0.35i}$. Combining these, we find total viscoelastic coefficients $\nu_{xz} = 0.0108e^{0.30i}$ and $\nu_{yz} = 0.0336e^{-0.33i}$. Inserting the net viscoelastic coefficients into the simplified laminar model forced response, given by equations \eqref{eq:viscous_response_A}-\eqref{eq:viscous_response_B}, gives very good agreement with the amplitudes and phases measured from the simulation.

Compared with the previous hydrodynamic results explored in Section \ref{subsec:hydro_reynolds_stress}, we see that our MHD run produces slightly lower magnitude viscoelastic coefficients, which are compatible with the larger equilibrium sloshing amplitudes found in this section. This might initially be surprising, since the presence of the MRI could be thought to drive greater stresses. However, unlike the horizontal MRI stress, which is dominated by the Maxwell term, the vertical viscoelastic coefficients are predominantly hydrodynamic in character. As we will explore more in Section \ref{subsec:forcing_amplitude}, the MRI actually acts to suppress and damp the emergence of the  elevator modes associated with the PI and therefore inhibits the vertical communication between sloshing layers.

\subsection{Variation with Forcing Amplitude}
\label{subsec:forcing_amplitude}

\begin{figure*}
    \centering
    \includegraphics[width=\linewidth]{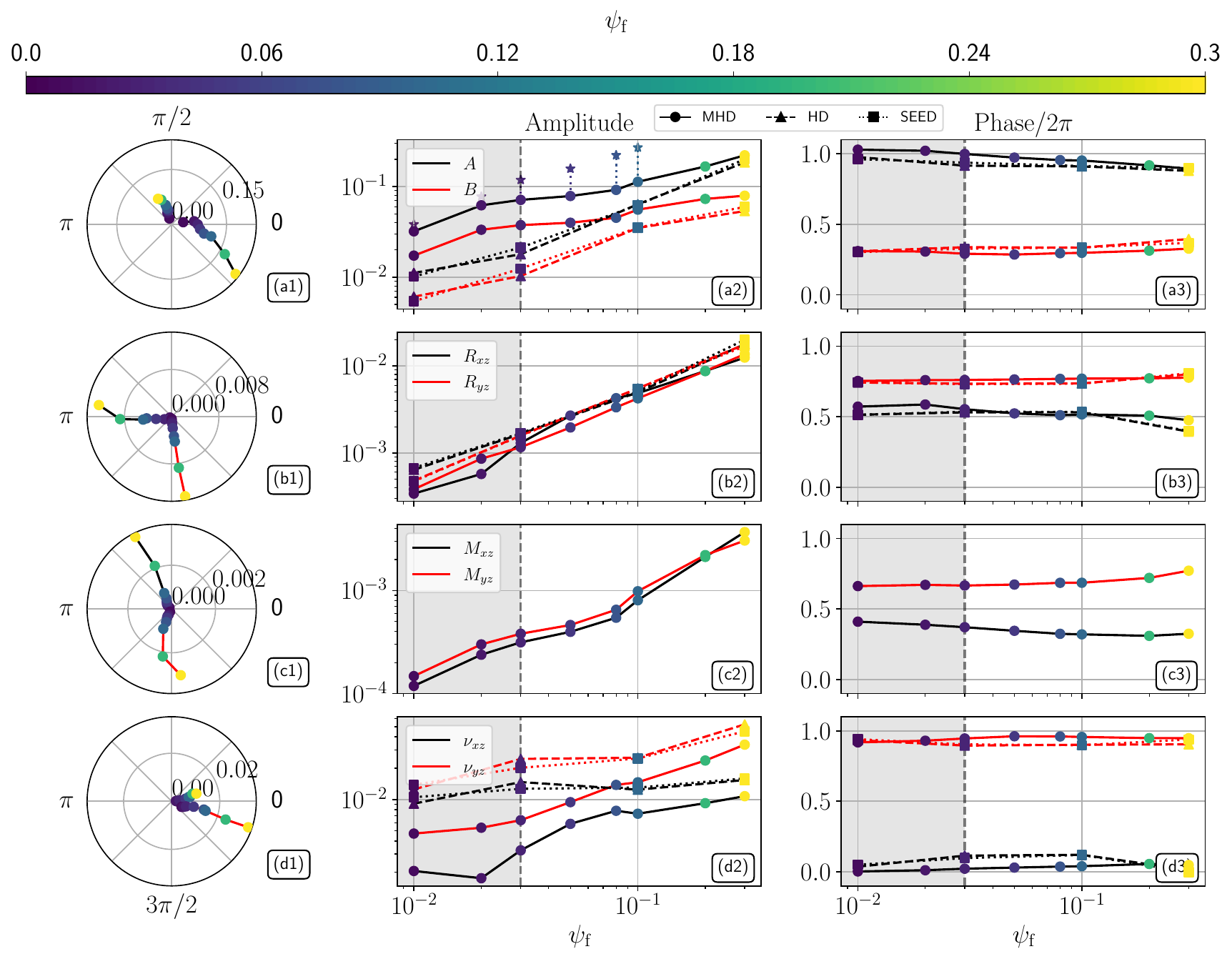}
    \caption{Variation in the sloshing diagnostics with $\psi_{\rm f}$. \textit{Row (a)}: Sloshing motion $A$ and $B$. \textit{Row (b)}: Reynolds stresses. \textit{Row (c)}: Maxwell stresses. \textit{Row (d)}: viscoelastic constants. Column (1) visualizes the full complex information in a polar plot, whilst columns (2) and (3) plot the amplitude and phase information, respectively, as a function of $\psi_{\rm f}$. The red and black line colors denote the $x$ vs $y$ variables, whilst the different line styles+markers denote the MHD, inviscid hydrodynamic and inviscid MHD-seeded hydrodynamic runs (see inset legend). The marker colors also denote the value of $\psi_{\rm f}$. Finally, the starred markers correspond to the maximum radial sloshing amplitude attained during the transient phase after the forcing is switched on. For further explanation, see the discussion in-text.}
    \label{fig:figure_psi_comparison}
\end{figure*}

Up until now, we have focused on the fiducial forcing amplitude with $\psi_{\rm f} = 0.3$. This value is sufficiently strong that the emerging sloshing equilibrium is well distinguished from the underlying MRI turbulence. However, in this section, we will consider how the sloshing response varies as a function of $\psi_{\rm f}$. To this end, we perform a suite of MHD simulations with the same setup as described in Section \ref{subsection:PI_MRI_growth} but with $\psi_{\rm f} \in \lbrace 0.01,0.02,0.03,0.05,0.08,0.1,0.2,0.3 \rbrace$. In all cases, we first allow the sloshing to progress through the initial transient growth before settling into a quasi-steady oscillation. We then monitor this between times $40 < t/T < 60$ to extract the sloshing amplitudes, stresses, and viscoelastic coefficients, as before. The results are shown in Fig.~\ref{fig:figure_psi_comparison}. Row (a) plots the extracted sloshing amplitudes, rows (b) and (c) show the Reynolds and Maxwell stresses, respectively, and row (d) shows the viscoelastic constants. Column (1) visualizes the full complex amplitude and phase information in the form of polar plots. Meanwhile, columns (2) and (3) break this up into the amplitude and phase variation as a function of $\psi_{\rm f}$. In all cases, the colored markers denote the different values for $\psi_{\rm f}$ whilst the solid, connective colored black and red lines with circular markers pertain to the `$x$' versus `$y$' quantities, as denoted by the inset labels. For comparison, we also plot a sample of purely hydrodynamic runs in columns (2) and (3). Repeating the methodology described in Section \ref{subsec:hydro_numerical_setup} for $\psi_{\rm f} \in \lbrace 0.01,0.03,0.1,0.3\rbrace$, we similarly extract the sloshing diagnostics between $40 < t/T < 60$ and plot the results as the dashed lines with triangular markers. Finally, we repeat these hydrodynamic runs but now start from the perturbed velocity and density initial condition inherited from the final snapshot of our base MRI run, with the magnetic field now switched off. These are shown as the dotted lines with square markers. It should be noted that for these hydrodynamic runs, the explicit viscosity is switched off in the simulations, in keeping with the setup described in Section \ref{sec:hydro_slosh}. One might reasonably ask whether the explicit viscosity in the MHD experiments, versus the lack thereof in the hydrodynamic runs, is the main culprit behind differences in the final sloshing state. Therefore, in appendix \ref{app:visc_hydro}, we repeat the hydrodynamic runs but with an explicit viscosity that matches our MHD runs. We find essentially the same behavior as our inviscid hydrodynamic runs across the range of forcings considered. This underlines that the main conclusions of this section are insensitive to the small values of $\nu$ and it is in fact the MRI turbulence that leads to the primary differences between the MHD and hydrodynamic results.

Across all runs, panel (a2) clearly shows the unsurprising increase in the sloshing response with the forcing amplitude. However, it should be noted that this increase is sublinear in character. Indeed, fitting a power law to the MHD data yields $|A| \propto \psi_{\rm f}^{0.51}$. This owes to the fact that the viscoelastic coefficients also increase with the forcing in panel (d2). For example, the MHD runs are fit by $|\nu_{xz}| \propto \psi_{\rm f}^{0.57}$. These scalings are therefore approximately consistent with the the simplified amplitude equilibrium described by equation \eqref{eq:simple_alpha}. Such fits to the viscoelastic coefficients, as a function of $\psi_{\rm f}$, therefore outline a predictive procedure wherein the sloshing response can be obtained via the laminar viscoelastic equations \eqref{eq:viscous_response_A}-\eqref{eq:viscous_response_B}.

Returning to broad trends, we see that for lower values of $\psi_{\rm f}$, the ratio of $|A|/|B| \sim 2$ whilst the phase difference in panel (a3) is also roughly $\pi/2$, in keeping with a free epicylic motion. As $\psi_{\rm f}$ increases, this ratio $|A|/|B|$ rises and the phase difference trends toward $\pi$. Comparing the various physical setups, we clearly observe differences between the corresponding line styles. Notably, the magnetized, MRI runs typically saturate with a higher sloshing amplitude compared with the hydrodynamic runs, particularly for intermediate values of $\psi_{\rm f}$. This is equivalently borne out by the magnetized runs showing lower viscoelastic amplitudes in panel (d2). This might seem surprising at first, since naively the action of the MRI might be thought to enhance the damping of the sloshing motions. However, the MRI turbulence also acts to damp the growth of the parametric instability and the associated elevator modes. Since the hydrodynamic, PI modes dominate the vertical stress, suppression of these via the MRI allows the sloshing to attain a larger amplitude. The fact that the hydrodynamic run, with the MRI inherited initial conditions, also deviates from the full MHD run, suggests that the final sloshing state is indeed sensitive to the different physics and not just the seeding of perturbations. 

In all panels, the vertical, dashed line and the bounded gray area, for $\psi_{\rm f} < \psi_c = 0.03$, delimit the region for which the MRI is found to suppress the growth of the PI in our MHD runs. Namely, below this value, the runs did not show the development of elevator modes and instead the turbulence resembled the MRI base state. Meanwhile, the elevator modes and PI are still found to emerge below this value in our purely hydrodynamic runs. This MHD suppression of the PI below $\psi_c$ also seems to manifest in the slopes of the sloshing and viscoelastic amplitudes. The sloshing amplitude in panel (a2) rises more quickly with $\psi_{\rm f}$ when the MRI suppresses the PI, before the `knee' at $\psi_c$, where the response flattens. The viscoelastic coefficients also seem to rise more rapidly beyond this critical forcing as the PI overcomes the MRI. 

\begin{figure}
    \centering
    \includegraphics[width=\linewidth]{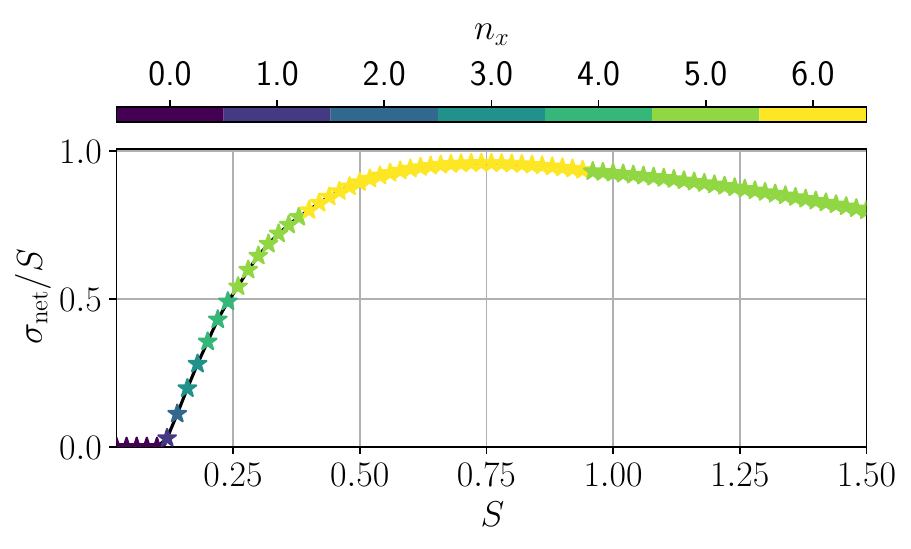}
    \caption{Model for the competition between PI growth and MRI damping of elevator modes. The maximum net growth rate $\sigma_{\rm net}$, normalized by the sloshing amplitude $S$, is plotted as a function of $S$. The color of the markers denotes the preferred radial order $n_x$ corresponding to this maximal growth rate.}
    \label{fig:figure_PI_MRI_toy}
\end{figure}

A simplified toy model might be invoked to motivate this critical forcing value. In particular, the elevator modes can emerge when the PI growth rate, at the maximal value of sloshing during the initial transient phase, exceeds the damping induced by the turbulent transport of momentum across the boundary between elevator mode channels. This turbulent damping of the elevator modes can be estimated as $\sigma_{\rm d}  \sim 4 k_x v_{x,{\rm MRI}}/(2\pi)$. The factor of 4 corresponds to the momentum transport in/out of the two boundaries, either side of a vertical elevator channel. The $k_x = 2\pi n_x/L_x$ encodes the radial width of the elevator mode, given the radial mode order $n_x$ fitting into the box of width $L_x$. Finally, the $v_{x,{\rm MRI}}$ denotes a characteristic radial velocity associated with the MRI eddies. Here, we take this to be equal to the standard deviation of the velocity fluctuation distribution, $v_{x,{\rm MRI}} = 0.085$,  as found in Fig.~\ref{fig:figure_MRI_state}. This competes with the PI growth rates $\sigma_{\rm PI}$, which can be found using the Floquet theory (using $N_{\rm max} = 30$), as described in Section \ref{subsec:floquet}. We solve for the maximally growing mode across a number of sloshing amplitudes $S$ and radial modes $n_x \leq 30$. Then, for each value of $S$, we find the maximum net growth rate $\sigma_{\rm net} = \sigma_{\rm PI}-\sigma_{\rm d}$ and the corresponding optimal modal number $n_x$. The results are shown in Fig.~\ref{fig:figure_PI_MRI_toy}, where the maximal growth rate is normalized by the value of $S$. 
Furthermore, the color of the starred data points indicates the modal number $n_x$ with the largest growth rate. For $S \leq 0.1$ the MRI-induced damping suppresses the PI growth for all modes. However, above this value, our model predicts growth at $n_x = 2$, before shorter wavelengths can overcome the damping at larger values of $S$. As one progresses to larger values of $S$, we see the preference for $n_x = 5$, then $n_x = 6$, in keeping with the elevator flows seen in our fiducial runs. As one progresses above $S\sim1$ we find the optimal $n_x$ decreases again, in accordance with the leftward drifting peaks seen in Fig.~\ref{fig:figure_high_slosh_floquet}.

To draw a comparison with the behavior seen in our MHD simulations, we return to Fig.~\ref{fig:figure_psi_comparison} and plot the maximum radial sloshing amplitude during the transient phase as starred markers in panel (a2). For $\psi_{\rm f} = 0.03$ this yields a maximal sloshing of $|A| = 0.12$, which should be susceptible to growth of the $n_x = 2$ elevator mode according to Fig.~\ref{fig:figure_PI_MRI_toy}. Upon examining the sloshing state for this run, we do indeed see the emergence of this very mode, marking the transition to the PI-dominated regime. Note that the maximal sloshing amplitude is always reduced substantially as the PI feeds off this shear energy and establishes stresses that equilibrate the sloshing at a lower amplitude.

%% file: 5_discussion.tex
\section{Discussion}
\label{sec:discussion}

This study builds upon previous efforts to understand the behavior of instabilities and turbulent stresses in the oscillatory flows established by distorted disks. Notably, \cite{TorkelssonEtAl_2000} also considered the interplay of the MRI and PI in stratified local box simulations. However, they treat the sloshing as a free initial condition, which rapidly decays, compared with our steady forcing setup. They attribute the initial rapid damping to the emergence of the PI, which is then followed by a period of exponential decay as the sloshing interacts with the magnetized turbulence. This allows them to indirectly infer an effective vertical $\alpha \sim 0.006$, which is comparable to the standard horizontal, base MRI stress they find -- broadly in support of an isotropic $\alpha$ prescription. However, whilst they measure horizontal stresses, which are dominated by the Maxwell component, they are surprised to find vertical stresses, which are primarily Reynolds. In fact, this is in keeping with our findings that the strong vertical flow structures set up by the PI survive and are primarily responsible for resisting the sloshing motion. Furthermore, our detailed analysis of the stress versus rate of strain relationship advocates for a more viscoelastic behavior, which is generally anisotropic between the $\nu_{xz}$ and $\nu_{yz}$ components. Because of the inherent epicylic motions sourced by the MRI, \cite{TorkelssonEtAl_2000} are fairly restricted to initializing large sloshing motions which stand out above the background. Whilst our fiducial runs also adopt this clear separation, the continual forcing permits us to track the equilibrated sloshing over many dynamical timescales and thereby extract a cleaner signal for small $\psi_{\rm f}$. This yields sloshing responses and viscoelastic amplitudes which vary as a complicated function of $\psi_{\rm f}$,  as seen in Fig.~\ref{fig:figure_psi_comparison}. Notably, we suggest that the MRI counterintuitively promotes larger sloshing flows as it tends to suppress the vertical hydrodynamic stresses that the PI wants to set up.

Another point of comparison is the purely hydrodynamic, two-dimensional setup considered by \cite{PaardekooperOgilvie_2019}, who followed the growth and saturation of the PI in the warped shearing box formalism of \cite{OgilvieLatter2013b}. This implementation fully accounts for the warped geometry and stratified disk structure, finding a richness of behavior. Notably, the PI can significantly suppress the laminar flows and reduce the efficacy of warp communication. Whilst we ramp up the sloshing from rest, they typically start from the steady laminar flow set by some explicit value of $\alpha$. For larger $\alpha$, they find that the PI growth is weaker and the effect on the laminar flows is more modest. This seems analogous to the presence of the MRI in our setup, which damps the strength of the emergent PI. They also briefly examine the resonant, inviscid limit for which no laminar steady state exists and instead ramp up the sloshing from rest, finding results consistent with their low $\alpha$ runs. The nature of the final sloshing state is found to depend sensitively on which PI radial modal number dominates. Just as we find the emergence of specific elevator modes, they also find that the sloshing state is characterized by preferred wavelengths. Note that these can change in time, since as the sloshing evolves, the modal growth rates vary -- hence different modes become preferred and compete for dominance. This can lead to state transitions in the flow and also complex hysteresis effects as the final equilibrium depends on the history of the sloshing. We also observe similar behavior as the strong sloshing initially selects a relatively short wavelength elevator mode (see Fig.~\ref{fig:figure_high_slosh_floquet} panel (a)), before these often merge and are subsumed by longer wavelength channels at late times. 

Whilst the work of \cite{PaardekooperOgilvie_2019} is ideally designed to measure the internal torques relevant to the evolution of warped disks; their two-dimensional simulations perhaps limit the character of the ensuing turbulence and stresses. Whilst our setup is less obviously comparable to the realistic warping scenario, our three-dimensional study, including magnetic fields, provides a complementary perspective that proffers more insight into the combined PI-MRI phenomenology. One major caveat is the lack of stratification in our model, which makes a direct quantitative link to the full stratified, warping regime tenuous. However, we might be tempted to draw a qualitative analogy between regimes when the values of $\psi$ and $\psi_{\rm f}$ predict viscous laminar flows of comparable magnitude. In particular, for a stratified warped disk, one predicts the sloshing flow induced at $1H$ to be $\sim \psi\,c/\alpha$ (see Eq. (38) of \cite{LodatoPringle_2007}). Meanwhile, manipulation of \eqref{eq:simple_alpha} yields a maximum sloshing amplitude in our model of $\sim \psi_{\rm f}\, c/(8\pi^2 \alpha)$. This motivates the correspondence $\psi \sim \psi_{\rm f}/(8\pi^2)$, for which the equilibrium internal flows and average shear rates are similar across both models. This suggests our fiducial upper limit of $\psi_{\rm f} = 0.3$, might relate to lower values of $\psi  \sim 0.004$. Even for these small warps (below typical values for $H/R$), the sloshing motion can become of sonic order owing to the Keplerian resonance. As one progresses to larger warp amplitudes $\psi\sim H/R$, nonlinear effects also come into play wherein the stratified disk can substantially depart from hydrostatic balance. In this case, the disk can enter a `bouncing' regime with extreme scale height variations \citep[e.g.][]{FairbairnOgilvie2021b}. This forcing of the disk scale height offers yet another source of modulation to the flow structure, which has recently been studied by \cite{HeldOgilvie_2024}. This vertical oscillation is also found to be susceptible to another form of the PI, which taps into the free energy associated with the breathing motion and can generate bending waves. The interplay of the PI modes, driven simultaneously by both the radial sloshing and bouncing motions, merits further investigation. 

To this end, in future work, we will pursue a stratified extension of this experiment, which will allow us to explore some of these neglected effects and map more directly onto the internal flows in realistic warped disks. In particular, upper disk boundaries will confine the elevator flows and presumably modify the sloshing equilibria and associated viscoelastic coefficients.

%% file: 6_conclusion.tex
\section{Conclusion}
\label{sec:conclusion}

Motivated by the radial, oscillatory sloshing flows found in warped astrophysical disks, in this paper, we have performed the first detailed numerical study examining the interplay of the parametric and magnetorotational instabilities in forced, time-dependent shear flows within a local, unstratified shearing box. In the weak and unforced sloshing regime, we verify the growth of inertial waves in our simulations, in agreement with the predictions of the three-mode coupling theory. We then resonantly drive these sloshing flows and find the resulting equilibrated states across a suite of hydrodynamic and magnetized runs, exploring a range of forcing amplitudes. In both regimes, the parametric instability is typically found to dominate the vertical transport of horizontal momentum and temper the sloshing motions. In our hydrodynamic runs, the resonantly forced flows initially drive large epicylic oscillations, which are unstable to counter-propagating vertical elevator modes, representing the strong sloshing outcome of the parametric instability. Above a critical warping amplitude, the same qualitative behavior emerges in magnetorotationally unstable runs, albeit with slightly increased sloshing amplitude as the MRI turbulence damps the growth of the elevator modes, which are then less efficient at resisting the forcing. Below this critical amplitude, the elevator modes are completely suppressed, and the sloshing motion is mediated primarily by the MRI. A detailed comparison of the stress and rate of strain profiles suggests that the vertical stresses are more accurately described by anisotropic, viscoelastic coefficients as opposed to the simplistic $\alpha$ description. Nonetheless, over the range of forcings considered, the extracted vertical viscoelastic coefficients are of similar amplitude to the horizontal $\alpha$ viscosity associated with the standard MRI, despite their physical characters being emphatically distinct. This work paves the way for future investigation into magnetized warps, where we will extend our setup to include stratification, allowing for a more direct connection with realistic distorted disks.

%% file: app_parametric_instability.tex
\section{Parametric Instability Theory}
\label{app:wn_parametric_instability}

In order to benchmark our numerical setup and its capacity to simulate oscillatory, shear flows, it is instructive to develop an analytical theory for the dynamics of weakly nonlinear sloshing. Indeed, it has long been recognized that time-dependent shear flows are susceptible to a three-mode-coupling, parametric instability \citep[][]{GammieEtAl_2000}, which will be developed in detail for our unique setup in this section. Here, we will broadly follow the analysis expounded in \cite{OgilvieLatter2013b} and  \cite{FairbairnOgilvie_2023}, which instead dealt with isothermal and polytropic, stratified disks.

\subsection{Governing Equations}
\label{app:subsec:governing_equations}

The hydrodynamic and axisymmetric limit of the unstratified shearing box equations \eqref{eq:continuity}-\eqref{eq:momentum} can be written as
\begin{eqnarray}
    \label{eq:ux_sbox}
    & D u_x-2\Omega_0 u_y = -\partial_x h +2 q \Omega_0^2 x, \\
    \label{eq:uy_sbox}
    & D u_y+2\Omega_0 u_x = 0.0 ,\\
    \label{eq:uz_sbox}
    & D u_z = -\partial_z h ,\\
    \label{eq:h_sbox}
    & D h = -c^2 \nabla\cdot \mathbf{u} ,
\end{eqnarray}
where we have introduced the material derivative, $D\equiv \partial_t+\mathbf{u}\cdot\nabla$, and our thermodynamic quantities are replaced with the pseudo-enthalpy variable $h = c^2 \ln\rho$. As described in Section \ref{sec:problem_setup}, these equations readily admit the solution
\begin{eqnarray}
    \label{eq:free_slosh}
    && u_{x,{\rm w}} = S c \sin(k_{\rm b} z)\cos(\Omega_0 t) , \\
    && u_{y,{\rm w}} = u_{\rm K}-\frac{1}{2}S c \sin(k_{\rm b} z)\sin(\Omega_0 t) , \\
    && u_{z,{\rm w}} = 0.0 ,
\end{eqnarray}
where $u_{\rm K} = -q \Omega_0 x$ is the standard, azimuthal shear flow and $h = h_0$ is a uniform constant. Meanwhile, the terms proportional to the sloshing amplitude $S$ encapsulate the vertically shearing epicylic motions, typical of warped disks. We now perturb this solution so that $\mathbf{u} = \mathbf{u}_{\rm w}+\mathbf{u}^\prime$ and $h = h_0+h^\prime$. Inserting this into equations \eqref{eq:ux_sbox}-\eqref{eq:h_sbox} and discarding any nonlinear products of primed quantities, the linearized equations become
\begin{eqnarray}
    \label{eq:ux_pert}
    && \partial_t u_x^\prime + S c \sin(k_{\rm b} z)\cos(\Omega_0 t)\partial_x u_x^\prime + S c k_{\rm b} \cos(k_{\rm b} z)\cos(\Omega_0 t) u_z^\prime -2\Omega_0 u_y^\prime = -\partial_x h^\prime \\
    \label{eq:uy_pert}
    && \partial_t u_y^\prime + S c \sin(k_{\rm b} z)\cos(\Omega_0 t)\partial_x u_y^\prime
    - (S c k_{\rm b}/2) \cos(k_{\rm b} z)\sin(\Omega_0 t) u_z^\prime + (\Omega_0/2)u_x^\prime = 0 \\
    \label{eq:uz_pert}
    && \partial_t u_z^\prime + S c \sin(k_{\rm b} z)\cos(\Omega_0 t)\partial_x u_z^\prime = -\partial_z h^\prime \\
    \label{eq:h_pert}
    && \partial_t h^\prime +S c \sin(k_{\rm b} z)\cos(\Omega_0 t)\partial_x h^\prime = -c^2 \left(\partial_x u_x^\prime+\partial_z u_z^\prime\right).
\end{eqnarray}
We now Fourier decompose the perturbations according to $X^\prime = \sum_{n = -\infty}^{\infty} \tilde{X}_n \exp{\left[i(k_x x+2\pi n z/L_z)\right]}$. Inserting this modal ansatz into equations \eqref{eq:ux_pert}-\eqref{eq:h_pert}, we exploit the orthogonality of complex exponentials to project onto evolutionary equations for each vertical order $n$. Upon nondimensionalizing the problem we arrive at 
\begin{eqnarray}
    \label{eq:non_dimensional_un}
    && \partial_\tau u_n - (K_x S/2)\cos(\tau) \left( u_{n+1}-u_{n-1}\right)
     +(S K_{\rm b}/2)\cos(\tau)\left(w_{n+1}+w_{n-1}\right)
     -2 v_n+i K_x h_n = 0 ,\\
    \label{eq:non_dimensional_vn}
    && \partial_\tau v_n - (K_x S/2)\cos(\tau) \left( v_{n+1}-v_{n-1}\right)
    -(S K_{\rm b}/4)\sin(\tau)\left(w_{n+1}+w_{n-1}\right)+u_n/2 = 0  ,\\
    \label{eq:non_dimensional_wn}
    && \partial_\tau w_n - (K_x S/2)\cos(\tau) \left( w_{n+1}-w_{n-1}\right)+i K_{\rm b} n h_n = 0 ,\\
    \label{eq:non_dimensional_hn}
    && \partial_\tau h_n - (K_x S/2)\cos(\tau) \left( h_{n+1}-h_{n-1}\right)
    + i K_x u_n+i K_{\rm b} n w_n = 0 ,
\end{eqnarray}
where $\mathbf{u}_n = \tilde{\mathbf{u}}_n/c$ and $h_n = \tilde{h}_n/c^2$. Furthermore, we exploit the characteristic time scale $\Omega_0^{-1}$ and the length scale $H = c/\Omega_0$, to define the dimensionless time coordinate $\tau = \Omega_0 t$ and wavenumbers $K_x = k_x H$ and $K_{\rm b} = k_{\rm b} H$.

\subsection{Weakly Nonlinear Asymptotics}
\label{app:subsec:wn_asymptotics}

We can make analytical progress by assuming that the magnitude of the sloshing motion is weak, such that $S \ll 1$. In other words, the sloshing flows should be sufficiently subsonic. Leveraging $S$ as an asymptotic parameter, we assume that the perturbations admit expansions of the form $X_n  = X_n^{(0)}+S X_n^{(1)}+\mathcal{O}(S^2)$. We also allow the problem to evolve over multiple timescales $(\tau_0,\tau_1) = (\tau, S\tau)$, such that time derivatives $\partial_\tau = \partial_0+S\partial_1$, where $\partial_i = \partial_{\tau_i}$. This allows us to track the perturbation dynamics over the fast orbital timescale, whilst also resolving the modulation of this behavior over longer, secular timescales. Inserting this expansion into equations \eqref{eq:non_dimensional_un}-\eqref{eq:non_dimensional_hn} and gathering terms of equal order in $S$ yields a hierarchy of equations to be solved sequentially.

\subsubsection{Leading Order: $\mathcal{O}(S^0)$}

At leading order the equations reduce to 
\begin{equation}
    \mathbf{L}_n\mathbf{V}_n^{(0)} = 0 ,
\end{equation}
where $\mathbf{V}_{n}^{(0)} = [u_n^{(0)},v_n^{(0)},w_n^{(0)},h_n^{(0)}]^T$ and the linear operator is given by
\begin{equation}
    \mathbf{L}_n = 
    \begin{pmatrix}
        \partial_0 & -2 & 0 & i K_x \\
        1/2 & \partial_0 & 0 & 0 \\
        0 & 0 & \partial_0 & i K_{\rm b} n \\
        i K_x & 0 & i K_{\rm b} n & \partial_0 
    \end{pmatrix}.
\end{equation}
This set of equations describes simple, decoupled inertial-acoustic waves for the system. Indeed, if we insert a temporal Fourier ansatz $\mathbf{V}_n^{(0)} \propto e^{-i \omega_0 \tau_0}$ and set the matrix determinant equal to zero, we arrive at the dispersion relation 
\begin{equation}
    \label{eq:dispersion_relation}
    (-\omega_0^2+K_{\rm b}^2 n^2)(-\omega_0^2+1)-K_x^2\omega_0^2 = 0 .
\end{equation}
This describes a family of frequency curves as functions of the radial wavenumber, each branch labeled by the discrete vertical mode number $n$. For visualization, these inertial branches are plotted in Fig.~\ref{fig:figure_dispersion_relation} as different colors. The blue branch denotes the nonoscillatory $n=0$ mode with eigenvectors satisfying $v_n^{(0)} = iK_x h_n^{(0)}/2$, $u_n^{(0)} = 0$. This steady-state mode corresponds to an adjusted geostrophic balance between the radial pressure (enthalpy) gradients and the Coriolis force, induced through the perturbed azimuthal velocity. Meanwhile, the vertical velocity can flow freely with arbitrary $w_n^{(0)}$. In fact, this linear mode is also a nonlinear solution to the full equations \eqref{eq:ux_sbox}-\eqref{eq:h_sbox}. Although it does not participate in the weakly nonlinear parametric instability, it is clearly emergent in our strong sloshing numerical experiments (see Section \ref{subsec:instability_growth}). The lowest frequency oscillatory branch labels the $n=1$ vertical mode, which incrementally increases toward higher $n$ as one jumps to higher frequency branches at a given $K_x$. The eigenmodes associated with the $n\geq1$ branches are described by 
\begin{equation}
    \label{eq:eigenvector}
    \mathbf{V}_n^{(0)}(\omega;K_x) = 
    \begin{bmatrix}
       & i \omega (\omega^2-K_{\rm b}^2 n^2) \\
       & \frac{1}{2}(\omega^2-K_{\rm b}^2 n^2) \\
       & i K_x K_{\rm b} n \omega \\
       & i K_x \omega^2
    \end{bmatrix}
    e^{-i\omega\tau_0}.
\end{equation}
Finally, the rows of the $\mathbf{L}_n$ operator can also be combined into a single, fourth-order differential operator acting on the enthalpy perturbations according to 
\begin{equation}
    \label{eq:inertial_operator}
    \mathcal{L}_n [h_n^{(0)}] = \left[(\partial_0^2+K_{\rm b}^2 n^2)(\partial_0^2+1)-K_x^2\partial_0^2\right]h_n^{(0)} = 0.
\end{equation}
This is is clearly related to the dispersion relation shown in equation \eqref{eq:dispersion_relation}.

\subsubsection{First Order: $\mathcal{O}(S^1)$}

At next order in $S$ we arrive at the set of equations
\begin{equation}
    \label{eq:first_order}
    \mathbf{L}_n \mathbf{V}_n^{(1)} = \mathbf{F}_n^{(1)} ,
\end{equation}
where the right-hand side forcing vector is given by 
\begin{eqnarray}
    \label{eq:rhs_forcing}
    & \mathbf{F}_n^{(1)} = -\partial_1 \mathbf{V}_n^{(0)}+
     \begin{bmatrix}
        & \frac{1}{2} K_x \cos{\tau_0}(u_{n+1}^{(0)}-u_{n-1}^{(0)})-\frac{1}{2} K_{\rm b} \cos\tau_0 (w_{n+1}^{(0)}+w_{n-1}^{(0)}) \\
        & \frac{1}{2} K_x \cos{\tau_0}(v_{n+1}^{(0)}-v_{n-1}^{(0)})+\frac{1}{4} K_{\rm b} \sin\tau_0 (w_{n+1}^{(0)}+w_{n-1}^{(0)}) \\
        & \frac{1}{2} K_x \cos\tau_0 (w_{n+1}^{(0)}-w_{n-1}^{(0)}) \\
        & \frac{1}{2} K_x \cos\tau_0 (h_{n+1}^{(0)}+h_{n-1}^{(0)})
    \end{bmatrix}.
\end{eqnarray}
The inertial wave operator $\mathbf{L}_n$ is inherited from leading order and now acts on the first-order perturbations. The wave modes from zeroth order in $S$ now manifest as forcing contributions on the right-hand side. Notably, this includes `off-diagonal' coupling terms since the product of the $n=1$ sloshing structure with $n^{\rm th}$ order waves, facilitates communication with the neighboring $n\pm1$ modes. 

Since the singular left-hand side operator obviously admits inertial waves as nontrivial solutions, the forcing vector on the right-hand side must be orthogonal to these solutions in some sense. More specifically, consider combining the rows of equation \eqref{eq:first_order} into differential operator form. This yields an equation of the form
\begin{equation}
    \label{eq:}
    \mathcal{L}_n[h_n^{(1)}] = F_n^{(1)}.
\end{equation}
Since the left-hand side is simply the inertial wave operator from zeroth order for the $(K_x,n)$ mode, the right-hand side must contain no terms proportional to $e^{-i\omega_0 \tau_0}$, which would otherwise resonantly force a singular, nonphysical response. If the forcing vector $\mathbf{F}_n^{(1)}$ contains such resonant contributions, which we denote $\mathbf{F}_{n,\omega_0}^{(1)} = [A,B,C,D]^\mathsf{T}$, then in order for these to cancel upon combination under the row operations, we must impose the solvability condition
\begin{equation}
    \label{eq:solvability_condition}
    K_x \omega_0 ( \omega_0 A +2 i B )+(\omega_0^2-1)(K_{\rm b} n C +\omega_0 D) = 0 .
\end{equation}
This solvability condition will lead to a coupling between the wave modes found at zeroth order. In particular, the relevant modal combination should involve a three-mode resonance, where the parent sloshing wave interacts with two inertial daughter modes, with dimensionless angular frequencies denoted $\omega_1$ and $\omega_2$ and vertical modal numbers labeled $n_1$ and $n_2$. Since the sloshing is horizontally homogeneous, the inertial modes should share the same $K_x$. Meanwhile, in order to overlap with the vertical and temporal structure of the sloshing, we require $\omega_2 - \omega_1 = 1$ and $n_2-n_1 = 1$. Such resonant couplings are shown by the connective lines between neighboring branches of the dispersion relation curves in Fig.~\ref{fig:figure_dispersion_relation}. Constructing a modal sum that satisfies these conditions yields
\begin{equation}
\label{eq:mode_combination}
    \mathbf{V}^{(0)} = \mathbf{U}_n^{(0)}+\mathbf{U}_{n+1}^{(0)},
\end{equation}
where
\begin{eqnarray}
    &&\mathbf{U}_n^{(0)} = \mathcal{A}_n(\tau_1) \mathbf{V}_n^{(0)}(\omega_0;K_x), \\
    &&\mathbf{U}_{n+1}^{(0)} = \mathcal{A}_{n+1}(\tau_1) \mathbf{V}_{n+1}^{(0)}(\omega_0+1;K_x).
\end{eqnarray}
This modal sum can be inserted into equation \eqref{eq:rhs_forcing} for $\mathbf{F}_n^{(1)}$ and $\mathbf{F}_{n+1}^{(1)}$. In each case, we extract the terms proportional to $e^{i\omega_0 \tau_0}$ and $ e^{i(\omega_0+1) \tau_0}$, which would resonate with the left-hand side operator, giving
\begin{eqnarray}
    &&\mathbf{F}_{n,\omega_0}^{(1)} = -\partial_1 \mathbf{U}_n^{(0)} +
        \frac{1}{4}
        \begin{bmatrix}
        K_x U_{n+1}^{(0)} - K_{\rm b} W_{n+1}^{(0)} \\
        K_x V_{n+1}^{(0)} - \frac{i}{2} K_{\rm b} W_{n+1}^{(0)} \\
        K_x W_{n+1}^{(0)} \\
        K_x h_{n+1}^{(0)} \\        
        \end{bmatrix}
\end{eqnarray}
and
\begin{eqnarray}
    &&\mathbf{F}_{n+1,\omega_0+1}^{(1)} = -\partial_1 \mathbf{U}_{n+1}^{(0)}
        -\frac{1}{4}
        \begin{bmatrix}
        K_x U_{n}^{(0)} + K_{\rm b} W_{n}^{(0)} \\
        K_x V_{n}^{(0)} - \frac{i}{2} K_{\rm b} W_{n}^{(0)} \\
        K_x W_{n}^{(0)} \\
        K_x h_{n}^{(0)} \\        
        \end{bmatrix}.
\end{eqnarray}
Inserting each of these resonant forcing vectors into the solvability condition given by equation \eqref{eq:solvability_condition} yields a pair of coupled equations
\begin{eqnarray}
    \label{eq:An_growth}
    && \frac{\partial \mathcal{A}_n}{\partial\tau_1} = C_1 \mathcal{A}_{n+1},\\
    \label{eq:Anp1_growth}
    && \frac{\partial \mathcal{A}_{n+1}}{\partial\tau_1} = C_2 \mathcal{A}_{n},
\end{eqnarray}
governing the modal evolution over the slow timescales. The coupling coefficients are given by
\begin{eqnarray}
    \label{eq:C1}
     & C_1 = \frac{K_x (2 \omega_0 +1) \left(-K_{\rm b}^2 \omega_0^2-K_{\rm b}^2 n^2-K_{\rm b}^2 n \omega_0^2-K_{\rm b}^2 n+\omega_0^4+2\omega_0^3+\omega_0^2\right)}{8\omega_0\left(\omega_0^4-K_{\rm b}^2 n^2\right)} \\
    \label{eq:C2}
     & C_2 = \frac{K_x (2 \omega_0 +1) \left(-K_{\rm b}^2 n^2+K_{\rm b}^2 n \omega_0  (\omega_0 +2)+\omega_0^2 (\omega_0 +1)^2\right)}{8 (\omega_0 +1) \left(K_{\rm b}^2 (n+1)^2-(\omega_0 +1)^4\right)}.
\end{eqnarray}
Combining equations \eqref{eq:An_growth}-\eqref{eq:Anp1_growth}, and converting back to the fast timescale yields the growth rate 
\begin{equation}
    \frac{\sigma}{\Omega_0} = S\sqrt{C_1 C_2}.
\end{equation}

%% file: app_visc_hydro.tex
\section{Viscous Hydrodynamic Runs}
\label{app:visc_hydro}

\begin{figure*}
    \centering
    \includegraphics[width=\linewidth]{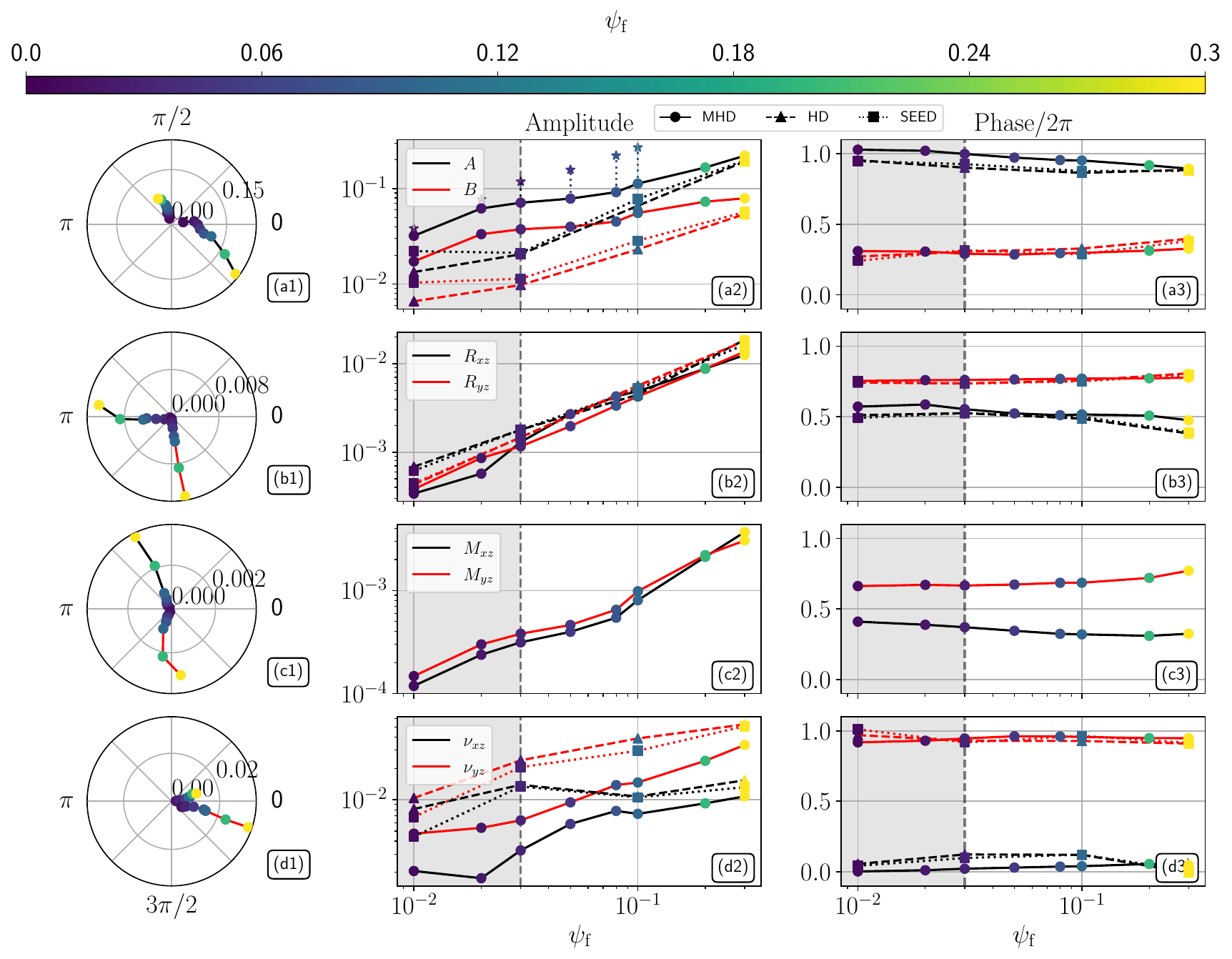}
    \caption{The same as Fig.~\ref{fig:figure_psi_comparison}, but now the hydrodynamic simulations include the explicit viscosity $\nu = 3.2\times10^{-4}$. These viscous hydrodynamic data points behave very similarly to the inviscid case, and preserve the main differences with respect to the MRI runs.}
    \label{fig:figure_psi_comparison_visc_hydro}
\end{figure*}

In sections \ref{sec:hydro_slosh} and \ref{sec:mhd_slosh}, we perform inviscid hydrodynamic simulations, setting the explicit viscosity $\nu = 0$. However, our MHD runs in Section \ref{sec:mhd_slosh} invoke a nonzero value of $\nu = 3.2\times10^{-4}$ in order to develop a converged MRI. One might be concerned that this small, explicit viscosity could be an important factor in controlling the sloshing response. Indeed, in the weak sloshing limit (see Section \ref{sec:parametric}), sufficient viscous damping would suppress the growth of the PI. Therefore, in order to check that this viscosity difference does not lead to the different sloshing behavior between the MHD and hydrodynamic cases, as seen in Fig.~\ref{fig:figure_psi_comparison}, we have repeated the hydrodynamic simulations with $\nu = 3.2\times10^{-4}$. These new viscous hydrodynamic results are shown in Fig.~\ref{fig:figure_psi_comparison_visc_hydro}, which otherwise has formatting equivalent to Fig.~\ref{fig:figure_psi_comparison}.

Comparison of Figs.~\ref{fig:figure_psi_comparison} and \ref{fig:figure_psi_comparison_visc_hydro} quickly demonstrates that the viscous and inviscid data points are nearly everywhere close to each other. More importantly, they remain clearly distinct from the trends set by the MHD runs. This indicates that the main differences between the hydrodynamic and MRI runs are not due to the small choice of explicit viscosity. Rather, the MRI turbulence is the fundamental disruptor that moderates the strength of the PI and elevator modes, as discussed in Section \ref{subsec:forcing_amplitude}. Indeed, we still clearly find that elevator modes emerge below the proposed critical value of $\psi_{c}\sim 0.3$ in these viscous hydrodynamic runs. The viscosity is sufficiently weak that the resonantly forced sloshing flows still grow enough during the early transient phase to be susceptible to the PI. Therefore, it must be the MRI, rather than viscosity, which suppresses the elevator modes in the weak forcing regime.